\documentclass[a4paper,11pt,oneside]{article}
             
\usepackage{jheppub} 
\usepackage[T1]{fontenc} 
\usepackage[utf8]{inputenc}
\usepackage{physics}
\usepackage{xcolor}
\usepackage{mathrsfs}
\usepackage{tikz}
\usepackage{colortbl}

\allowdisplaybreaks

\usepackage[framemethod=TikZ]{mdframed}
\mdfsetup{roundcorner=5pt}
   
\newmdenv[
skipabove=5pt,
skipbelow=7pt,
rightline=false,
leftline=false,
topline=false,
bottomline=false,
backgroundcolor=gray!30,
innerleftmargin=5pt,
innerrightmargin=5pt,
innertopmargin=5pt,
innerbottommargin=5pt,
leftmargin=0cm,
rightmargin=0cm,
linewidth=4pt]{eBox}

\hypersetup{colorlinks=true,urlcolor=pacificblue,citecolor=magenta,linkcolor=blue}

\graphicspath{{figures/}}

\definecolor{pacificblue}{cmyk}{0.95,0.3,0, 0.5}
\definecolor{Blue}{HTML}{2171b5}

\newcommand{\nc}{\newcommand}
\nc{\ba}{\begin{eqnarray}}
\nc{\ea}{\end{eqnarray}}

\newcommand{\calP}{{\cal{P}}}
\newcommand{\calH}{{\cal{H}}}
\nc{\bfx}{{\bf{x}}}
\nc{\bfk}{{\bf{k}}}
\nc{\mhn}{\color{blue}{\bf MHN: }}

\title{\fontfamily{lmr}\selectfont{Effective theories for a nonrelativistic field in an expanding universe: Induced self-interaction, pressure, sound speed, and viscosity}}

\author[a]{ Borna Salehian,}
\author[a]{Mohammad Hossein Namjoo,}
\author[b]{and David I. Kaiser}
\affiliation[a]{School of Astronomy, Institute for Research in Fundamental Sciences (IPM), Tehran, Iran}
\affiliation[b]{Department of Phyics, Massachusetts Institute of Technology, Cambridge, Massachusetts, USA}
\emailAdd{salehian@ipm.ir}
\emailAdd{mh.namjoo@ipm.ir}
\emailAdd{dikaiser@mit.edu}

\abstract{A massive, nonrelativistic scalar field in an expanding spacetime
is usually approximated by a pressureless perfect fluid, which leads to the standard conclusion that such a field can play the role of cold dark matter. In this paper, we systematically study these approximations, incorporating subleading corrections. We provide two equivalent effective descriptions of the system, each of which offers its own advantages and insights: (i) A nonrelativistic effective field theory (EFT) with which we show that the relativistic corrections induce an 
effective self-interaction for the nonrelativistic field. As a byproduct, our EFT also allows one to construct the exact solution, including oscillatory behavior, which 
is often difficult to achieve from the exact equations. (ii) An effective (imperfect) fluid description, with which we demonstrate that, for a perturbed Friedmann-Lema\^{i}tre-Robertson-Walker (FLRW) universe: (a) The pressure is small but nonzero (and positive), even for a free theory with no tree-level self-interactions. (b) The sound speed of small fluctuations is also nonzero (and positive), reproducing already known leading-order results, correcting a subdominant term, and identifying a new contribution that had been omitted in previous analyses. (c) The fluctuations experience a negative effective bulk viscosity. The positive sound speed and the negative bulk viscosity act in favor of and against the growth of overdensities, respectively. The net effect may be considered a smoking gun for ultra-light dark matter. }

\newcommand{\eq}[1]{\begin{equation}#1\end{equation}}
\newcommand{\eqa}[1]{\begin{align}#1\end{align}}
\newcommand{\spl}[1]{\begin{split} #1 \end{split}}
\newcommand{\fg}[1]{\begin{figure}[tbp]\centering #1 \end{figure}}

\newcommand{\p}{\partial}
\newcommand{\dpsi}{\delta\psi}
\newcommand{\bpsi}{\bar{\psi}}
\newcommand{\dro}{\delta\rho}
\newcommand{\dP}{\delta p}
\newcommand{\du}{\delta u}
\newcommand\Mpl{{M_{\rm Pl}}}
\newcommand{\cs}{c_s^2}
\newcommand{\cpsi}{\tilde{\psi}}
\newcommand{\cH}{\tilde{H}}
\newcommand{\cdpsi}{\tilde{\delta\psi}}
\newcommand{\cnabla}{\tilde{\nabla}}

\begin{document}
\maketitle
\flushbottom

\vspace{0.5cm}
\newpage
\section{Introduction}\label{sec:intro}
Ultra-light particles have been considered a promising candidate for dark matter \cite{Hu:2000ke,Hui:2016ltb,Ferreira:2020fam}. Axions, axion-like particles, and fuzzy dark matter are examples of this class of dark matter models. They arise naturally in UV-complete theories like string theory \cite{Arvanitaki:2009fg}. Presumably any such dark-matter candidates would arise in a fully relativistic setting, though their astrophysical implications would become most relevant deep in the nonrelativistic regime. Hence it is imperative to develop a self-consistent, effective description of the low-energy dynamics of such particles in an expanding spacetime, which incorporates corrections from subleading contributions. Building on Refs.~\cite{Guth:2014hsa,Namjoo:2017nia}, we develop such a formalism here. Our analysis applies to any massive scalar field oscillating in an expanding universe, for which interactions with relativistic fields can be neglected, including cases of physical interest such as ultra-light dark matter.

In this paper we study the dynamics of a massive scalar field $\phi$ minimally coupled to gravity. Our main interest concerns the case in which such a field plays the role of dark matter, although our analysis applies to other situations as well. The particles being ultra-light implies that, to account for the observed density of dark matter, the occupation numbers of $\phi$ particles must be very large. This, in turn, suggests that the system should be describable by a classical scalar field to a good approximation \cite{Guth:2014hsa,Hertzberg:2016tal}. On the other hand, the dark matter is supposed to be cold, and therefore nonrelativistic. Putting these two statements together, we deduce that a {\it classical field theory in the nonrelativistic limit} should provide an appropriate description of such a system.

The dynamics of a classical scalar field of mass $m$ in the nonrelativistic limit should be dominated by oscillations with nearly constant frequency $m$. That is, the energy of $\phi$ should be dominated by its rest mass. Although the scalar field oscillates very fast (compared to the Hubble scale, for example), we develop a formalism with which to study the dynamics of the system over time-scales much longer than $1/m$. If we introduce the ansatz \cite{Guth:2014hsa}
\eq{
	\label{eq:anphi}
	\phi (t, {\bf x} )= \dfrac{1}{\sqrt{2m}}\left( e^{-imt}\psi (t, {\bf x} ) +e^{imt}\psi^* (t, {\bf x} ) \right)\,,	
}   
where $\psi$ is a slowly varying function of time, then (as we will demonstrate below) the energy density of $\phi$ in a Friedmann-Lem\^{a}itre-Robertson-Walker (FLRW) universe is given (to leading order) by
\eq{
	\label{eq:endens}
	\rho= m|\psi|^2\propto a^{-3}\,,
}
which is slowly varying compared to the frequency of oscillations $m$. On the other hand, to leading order the pressure only contains rapidly oscillating terms which average out to zero on time scales larger than $1/m$. 

This motivates an approach that focuses on the slowly varying variable $\psi$.  We therefore take the relation \eqref{eq:anphi} as a field redefinition, with which to define the {\it nonrelativistic field} $\psi$. We will consider this field redefinition in more detail in the next section. Note that the $\psi$ field will not remain entirely as a slowly varying function of time; subdominant but highly oscillating contributions will develop as a result of nonlinear dynamics of the system. Because we are interested in the slowly varying part of the field, we develop techniques with which to remove the rapidly oscillating terms systematically. The result is an effective field theory (EFT) for the slowly varying part. The main aim of this paper is to derive such an EFT. 

Our nonrelativistic EFT will be an expansion in powers of several small parameters. To identify the relevant small parameters, 
 consider a massive, possibly self-interacting, real-valued scalar field $\phi$ coupled minimally to gravity, with action given by
\eq{
	\label{eq:action}
	S=\int\dd[4]{x}\left[\frac{1}{2}\Mpl^2\sqrt{-g}R+\mathcal{L}_\phi\right]\,,
}
where the matter Lagrangian is
\eq{
	\label{eq:lagphi}
	\mathcal{L}_\phi=-\frac{\sqrt{-g}}{2}\left[g^{\mu\nu}\p_\mu\phi\p_\nu\phi+m^2 \phi^2+\dfrac{\lambda}{4!}\phi^4\right]\,.
}
The resulting Klein-Gordon equation in an expanding universe reads
\ba 
\ddot \phi +3H\dot \phi -\dfrac{\nabla^2}{a^2}\phi+m^2 \phi +\dfrac{\lambda}{6}\phi^3=0 .
\label{eq:KleinGordon}
\ea 
 In the nonrelativistic limit, any dimensionful parameter with dimension mass should be smaller than the mass of the scalar field. Given the form of Eq.~(\ref{eq:KleinGordon}), this suggests three relevant, small quantities: 
\ba 
\label{eq:eps}
\epsilon_{\scriptscriptstyle H} \sim \dfrac{H}{m}\, , \qquad \epsilon_k \sim \dfrac{k_p^2}{m^2} \, , \qquad \epsilon_\lambda \sim \lambda \dfrac{\phi^2}{m^2},
\ea 
where $k_{p}$ is a typical physical wavenumber. Furthermore, after removing $\phi$ in favor of the new field $\psi$ as in Eq.~\eqref{eq:anphi}, all remaining variables are expected to be slowly varying compared to the time-scale $1/m$. This suggests a fourth small parameter, 
\ba 
\epsilon_t \sim \dfrac{1}{mX}\dfrac{\partial X}{\partial t},
\label{eq:ept}
\ea 
where $X$ is a variable in the theory. We will make our expansion in terms of such small parameters more explicit in the following sections. We will see that this perturbative expansion also allows us to construct the full solution, including oscillations, by solving well-behaved non-oscillatory equations. This can be considered a reliable method for finding solutions of an otherwise generically stiff system of equations with rapidly oscillating terms, with much broader application than what shall be discussed here.

In this paper we mainly focus on the case of a free, massive scalar field, setting $\lambda = 0$. As we will see, even in the absence of tree-level self-interactions for $\phi$, the backreaction from rapidly oscillating modes on the evolution of the nonrelativistic system in an expanding FLRW spacetime induces subdominant self-interaction terms in the evolution of $\psi$. The induced self-coupling could be of interest in the light of astrophysical observations \cite{Bullock:2017xww} which suggest that cold dark matter particles might depart from a purely collisionless regime \cite{Fan:2016rda,Tulin:2017ara,Ferreira:2020fam}, though a direct comparison of effects arising from the induced self-interaction with recent astrophysical observations remains beyond the scope of the present paper.
In Appendix~\ref{app:interaction} we consider the case of tree-level self-interactions as well.

Besides deriving an EFT for the nonrelativistic field, we also develop an equivalent description of the system in terms of an imperfect fluid. We show that in the latter description, one needs to identify a nonzero pressure, a nontrivial sound speed, and an effective bulk viscosity to fully express the EFT, up to the working order, in the fluid language. 

While the effect of oscillations of ultra-light particles as dark matter are neglected in a majority of related analyses, they are studied in a rather restricted literature.  The effect of oscillations on the dynamics of binaries is studied in Ref.~\cite{Blas:2019hxz}, while Ref.~\cite{Boskovic:2018rub} explores their impact on the orbital motion of stars in galactic halos. The resulting parametric resonance are also studied e.g. in Ref.~\cite{Tkachev:1986tr} and recently in Ref.~\cite{Hertzberg:2020dbk}. Ref.~\cite{Cookmeyer:2019rna} also compares the predictions for effects on the cosmic microwave background (CMB) from the naive theory, in which oscillations are neglected, with predictions from the exact theory.

Let us make some remarks on terminology used in this paper. We work up to linear order in spatially varying fluctuations around a homogeneous and isotropic background. We are interested in nonlinearities in the time evolution of various quantities which occur even at this order; these nonlinearities arise because of the coupling between different time-dependent variables, which yield a {\it mode coupling} once we decompose the relevant quantities into series of modes with distinct characteristic time dependencies. To avoid confusion and simplify the terminology, we refer to quantities that vary in space (around a homogeneous, isotropic background) as ``fluctuations,'' and restrict ``(non)linearities'' to features of the time evolution of various quantities. Furthermore, as noted above, we are interested in an EFT for the slowly varying parts of the fields, which we denote the ``slow modes,'' while we systematically remove the rapidly oscillating contributions, which correspond to nonzero modes in the mode decomposition (in time) of various quantities. Finally, we adopt a perturbative scheme, accounting for various contributions to the effective equations of motion in terms of different powers of small parameters. We therefore distinguish between ``perturbations'' (as powers of these small expansion parameters) and ``fluctuations'' (as spatially varying quantities around the FLRW background).

The rest of the paper is organized as follows. In Sec.~\ref{sec:redef} we introduce a suitable field redefinition for the scalar field in a general spacetime geometry as well as in a perturbed FLRW universe. In Sec.~\ref{sec:EFT}, we derive our EFT for the background evolution as well as fluctuations. In Sec.~\ref{sec:fluid} we re-express our EFT results in terms of fluid dynamics. In Sec.~\ref{sec:num} we confirm the validity of our results by numerical analysis, and in Sec.~\ref{sec:con} we summarize our main results and discuss possible future directions. Furthermore, in a number of appendices, we explore different corners of the subject and extend the analyses presented in the main body of the paper. In Appendix~\ref{app:Hamiltonian} we derive the Hamiltonian of the system after the field redefinition discussed in Sec.~\ref{sec:redef}. In Appendix~\ref{app:linhigher} we present the equations for higher-order nonzero modes, ignored in the results of Sec.~\ref{sec:EFT}. In Appendix~\ref{app:vis} we discuss how an imperfect fluid can be described in the general relativistic framework, the results of which are used for the fluid interpretation of our EFT in Sec.~\ref{sec:fluid}. In Appendices~\ref{app:interaction} and \ref{app:multi-component}, we extend our results to the self-interacting case and to a multicomponent universe, respectively. In Appendix~\ref{app:nonlocal} we employ a nonlocal field redefinition in Minkowski spacetime to confirm the higher-order, momentum-dependent terms we obtained in the effective sound speed (which has a different numerical factor compared to previous analyses). The resulting sound speed is valid for arbitrary momentum and reduces to the well-known results in the limits of small and large momenta. Finally, in Appendix~\ref{app:gauge} we derive the gauge transformation that can be used to express our EFT (written in Newtonian gauge) in other gauges.

\section{A suitable field redefinition}\label{sec:redef}
Taking the nonrelativistic limit of a scalar field theory is usually done by starting with an appropriate field redefinition, which is the subject of this section. Similar to the approach in Ref.~\cite{Namjoo:2017nia}, we aim to take the nonrelativistic limit and obtain higher-order corrections to a low-energy EFT {\it after} performing the field redefinition. 

We are interested in studying the system at low energies, for which the  mass term is the dominant contribution in the system's evolution. That is, we expect the main time dependence of the system to be rapid oscillations with frequency approximately equal to the mass $m$. This motivates a natural choice for a complex-valued nonrelativistic field, $\psi$, related to $\phi$ as in Eq.~(\ref{eq:anphi}). The prefactor $1/\sqrt{2m}$ in Eq.~(\ref{eq:anphi}) has been chosen so that $|\psi|^2$ corresponds to the number density of particles, as will become evident below. We will also see that this field transformation, to leading order, yields a Schr\"odinger-like equation for $\psi$ in Minkowski spacetime.

As discussed above, $\psi$ is expected to be a slowly varying function of time in the nonrelativistic limit. Note that since, by Einstein's equations, the metric components are sourced by the energy-momentum tensor, which is quadratic in the original field $\phi$ --- and hence in the new field $\psi$ --- the metric components are dominated by slowly varying functions of time, although they also contain subdominant oscillatory contributions. For this reason, we do not make a similar field redefinition for the spacetime metric. 

As mentioned in Ref.~\cite{Namjoo:2017nia}, the field redefinition of Eq.~\eqref{eq:anphi} may not be considered complete if the real and imaginary parts of $\psi$ are treated as two independent variables, since that would consist of replacing one degree of freedom with two degrees of freedom. In Ref.~\cite{Namjoo:2017nia}, this is addressed by adding to Eq.~\eqref{eq:anphi} a comparable field redefinition for $\pi$, the conjugate momentum of $\phi$, such that the transformation from $(\phi,\pi)$ to $(\psi,\psi^*)$ becomes a canonical transformation. In addition, in Ref.~\cite{Namjoo:2017nia} a nontrivial nonlocal operator is used within the transformation, which simplifies the equation of motion for $\psi$ in Minkowski spacetime. In principle, a similar procedure can be employed when studying the behavior of $\psi$ in a curved spacetime. However, as we discuss in Appendix~\ref{app:nonlocal}, a nonlocal field redefinition does not in general result in any simplification in curved spacetime, even for the simple case of an exact FLRW universe. Notice that the nonlocal field redefinition is not essential for obtaining the nonrelativistic limit and we will not exploit it in a curved geometry.\footnote{See, however, Ref.~\cite{Friedrich:2019zic} where a generalization of the nonlocal operator introduced in Ref.~\cite{Namjoo:2017nia} to the case of expanding background has been employed.} Furthermore, making a transformation for $\pi$ similar to what was used in Ref.~\cite{Namjoo:2017nia} results in equations of motion different from the Schr\"odinger equation written in an FLRW background \cite{Guth:2014hsa}. Finally, we note that in terms of the original field $\phi$, the conjugate momentum differs from $\dot \phi$ in a curved spacetime, leaving more freedom for an appropriate field redefinition. 

Given these considerations, in this paper we follow a different approach compared to Ref.~\cite{Namjoo:2017nia}. From Eq.~\eqref{eq:anphi}, it is evident that there is a redundancy in the definition $\psi$ of the form
\eq{
	\label{eq:gaugetr}
	\psi (t, {\bf x}) \to\psi (t, {\bf x}) +ie^{imt}\eta (t, {\bf x}) \,,
}   
with $\eta$ any real function of space and time. It is easy to show that this transformation leaves $\phi$ --- and, as a result, the Lagrangian --- invariant. To fix the gauge and remove the redundancy, the following transformation from an arbitrary gauge seem to be appropriate:
\eq{
	\label{eq:gaugefix}
	\eta=\frac{1}{2m}\left(e^{-imt}\dot{\psi}+e^{imt}\dot{\psi}^*\right)\,.
} 
Note that this choice of the gauge guarantees that the following constraint is satisfied in the new gauge:
\eq{
\label{eq:gaugecon}
e^{-imt}\dot{\psi}+e^{imt}\dot{\psi}^*=0\,,
}
which allows us to have a first order, Schr\"odinger-like equation for $\psi$. Therefore, we may obtain the equations of motion for the new fields after gauge fixing by varying the following Lagrangian with respect to $\psi$ and $\psi^*$:
\eq{
\label{eq:lagmultiplier}
\mathcal{L}_\phi(\psi,\psi^*)+\xi (e^{-imt}\dot{\psi}+e^{imt}\dot{\psi}^*)\,,
} 
where $\mathcal{L}_\phi(\psi,\psi^*)$ is the Lagrangian given in Eq.~\eqref{eq:lagphi} with $\phi$ replaced by the field redefinition of Eq.~\eqref{eq:anphi}, and $\xi$ is a Lagrange multiplier, which ensures that the constraint of Eq.~(\ref{eq:gaugecon}) is satisfied as a result of gauge fixing. Note that in this gauge, we have
\eq{
	\label{eq:phidot}
	\dot{\phi}=-i\sqrt{\frac{m}{2}}\left(e^{-imt}\psi-e^{imt}\psi^*\right)\,,
}
or, by using Eqs.~\eqref{eq:anphi} and \eqref{eq:phidot}, 
\eq{
\label{eq:psiphi}
	\psi=\sqrt{\frac{m}{2}}e^{imt}\left(\phi+i\frac{\dot{\phi}}{m}\right)\,,
}
which shows that our field redefinition is invertible. Notice that the equation of motion for $\xi$ is a differential equation, rather than an algebraic constraint, and it is not possible to substitute the solution of $\xi$ back into the Lagrangian. In other words, the gauge condition cannot be imposed at the level of the Lagrangian. This makes the Lagrangian rather complex.
It is, however, possible to derive a different, and simpler, Lagrangian which yields the same equations of motion via the following procedure. First, in the original theory in terms of $\phi$, we introduce a new field by the relation $\chi = \dot \phi$ (at the level of equations of motion); next we replace $\dot \phi$ with $\chi$ in the Lagrangian and add a Lagrange multiplier to guarantee the equivalence of the two theories: 
\eq{
\label{eq:barL}
\bar{\mathcal{L}}_\phi(\phi,\chi)=\mathcal{L}_\phi(\phi,\chi,\p_i\phi)+\pi(\dot{\phi}-\chi)\,.
}
Note that at this stage, the field $\pi$ is only a Lagrange multiplier and \emph{not} the momentum conjugate of $\phi$. The equations of motion for this Lagrangian are as follows:
\eq{
	\label{eq:chiphi}
	\dot{\phi}-\chi=0\,,\qquad\pdv{\mathcal{L}_\phi}{\chi}-\pi=0\,,\qquad\pdv{\mathcal{L}_\phi}{\phi}-\p_i\pdv{\mathcal{L}_\phi}{\p_i\phi}-\dot{\pi}=0\,.
}
The first equation is the constraint. The second equation, together with $\chi=\dot{\phi}$, shows that the Lagrange multiplier is nothing but the momentum conjugate of the scalar field in the original Lagrangian (hence our motivation for the notation, $\pi$). Finally, using the first two equations, the last one gives the correct equation of motion for $\phi$. Therefore, the two theories $\mathcal{L}_\phi$ and $\bar{\mathcal{L}}_\phi$ are equivalent. 

In the theory governed by $\bar{\mathcal{L}}_\phi$, the parameter $\pi$ is nondynamical, so one can replace it by its solution to simplify the Lagrangian. From Eq.~\eqref{eq:chiphi} we have:
\eq{
\pi=	-\sqrt{-g}(g^{00}\chi+g^{0i}\p_i\phi)\,.
}
Plugging this solution back into the Lagrangian results in
\eq{
	\label{eq:phichilag}
\mathcal{L}(\phi,\chi)=-\frac{1}{2}\sqrt{-g}\left[-g^{00}\chi^2+2(g^{00}\chi+g^{0i}\p_i\phi)\dot{\phi}+g^{ij}\p_i\phi\p_j\phi+m^2\phi^2\right]\, ,
} 
where, from now on, we consider a non-self-interacting theory, i.e. we set $\lambda=0$ in Eq.~\eqref{eq:lagphi}. (We will consider tree-level self-interactions in Appendix~\ref{app:interaction}.)
Again, one can show that the Lagrangian in Eq.~(\ref{eq:phichilag}) yields consistent equations of motion. The next step is to introduce the nonrelativistic fields $\psi$ and $\psi^*$, which we define by the following expressions:
\eqa{
\label{eq:cantr1}
\phi=\frac{1}{\sqrt{2m}}\left(e^{-imt}\psi+e^{imt}\psi^*\right) \, , \qquad 
\chi=-i\sqrt{\frac{m}{2}}\left(e^{-imt}\psi-e^{imt}\psi^*\right)\,,
}  
which are indeed suggested by Eqs.~\eqref{eq:anphi} and \eqref{eq:phidot}. With these definitions, the Lagrangian in Eq.~\eqref{eq:phichilag} may be written
\eq{
\spl{
\mathcal{L}(\psi,\psi^*)=-\frac{\sqrt{-g}}{2}\Bigg[ &ig^{0\mu}(\psi^*\p_\mu\psi-\psi\p_\mu\psi^*)+\frac{1}{m}g^{\mu'\nu'}\p_{\mu'}\psi\p_{\nu'}\psi^*+m(g^{00}+1)\psi^*\psi\\
&+\left\{\frac{e^{-2imt}}{2m}\bigg(g^{\mu'\nu'}\p_{\mu'}\psi\p_{\nu'}\psi+\left(m^2(g^{00}+1)+\frac{im}{\sqrt{-g}}\p_\mu(\sqrt{-g}g^{0\mu})\right)\psi^2\bigg)+\text{c.c.}\right\}\Bigg]\, .
}
\label{eq:Lpsipsistar}
}
Here ``$\text{c.c.}$'' denotes the complex conjugate of the immediately preceding expression, and the primes on indices in $g^{\mu'\nu'}$ indicate that $\mu$ and $\nu$ cannot both be time components, i.e., $\mu\times \nu\neq0$. (In Appendix~\ref{app:Hamiltonian}, we consider the Hamiltonian of this theory in terms of $\psi$ and $\psi^*$.) From Eq.~(\ref{eq:Lpsipsistar}), the equation of motion for $\psi$ can be written as
\eq{
ig^{00}\dot{\psi}+\mathcal{D}\psi+e^{2imt}\mathcal{D}^*\psi^*=0\,,
\label{eq:psidotD}
}
where we have defined the operator
\eq{
\label{eq:D}
\mathcal{D}=\frac{m}{2}\left(g^{00}+1\right)+ig^{0i}\p_i+\frac{i}{2\sqrt{-g}}\p_\mu(\sqrt{-g}g^{0\mu})-\frac{1}{2m\sqrt{-g}}\p_\mu(\sqrt{-g}g^{\mu i})\p_i-\frac{1}{2m}g^{ij}\p_i\p_j\,.
}
As promised, the equation of motion for $\psi$ is first order in time derivatives and is identical to the equation of motion from the constrained Lagrangian of  Eq.~\eqref{eq:lagmultiplier}. In Minkowski spacetime the Lagrangian reduces to 
\ba 
\mathcal{L}(\psi,\psi^*)= \dfrac{i}{2}(\psi\dot \psi^*-\psi^*\dot \psi)-\frac{1}{2m}\p_{i}\psi\p_{j}\psi^*
-\left\{\frac{e^{-2imt}}{4m}\p_{i}\psi\p_{j}\psi+\text{c.c.}\right\}
 \, \, \text{for Minkowski}.\quad 
\ea 
Neglecting the last term, which is rapidly oscillatory, the resulting theory has a global $U(1)$ symmetry associated to which there is a conserved charge, $\int d^3 x|\psi|^2$, which is the number of particles, a quantity that is conserved in a nonrelativistic theory. The equation of motion reads
\eq{
i\dot{\psi}+\dfrac{\nabla^2}{2m}(\psi+e^{2imt}\psi^*)=0  \qquad \text{for Minkowski},
}
which, again after neglecting the oscillating term, is just the Schr\"odinger equation with vanishing potential. Note, however, that so far our equations are exact and only a field redefinition has been performed. We will see how one can remove the oscillatory terms in a systematic way, rather than just neglecting them, in Sec.~\ref{sec:EFT}.

In a curved background, the complete set of equations also includes the Einstein equations, $\Mpl^2G_{\mu\nu}=T_{\mu\nu}$, where the energy-momentum tensor for the scalar field is
\eq{
	T_{\mu\nu}=\p_\mu\phi\p_\nu\phi-\frac{1}{2}g_{\mu\nu}\left(g^{\alpha\beta}\p_\alpha\phi\p_\beta\phi+m^2\phi^2\right)\,,
	\label{eq:Tmn}
}
in which we must insert definitions of Eqs.~\eqref{eq:anphi} and \eqref{eq:phidot}. To avoid clutter we do not write out $T_{\mu\nu} (\psi, \psi^*)$ explicitly here.

\subsection{FLRW background}
What has been derived so far is valid for a general curved geometry. For the remainder of this paper, we consider the case of a spatially flat FLRW spacetime with small fluctuations around it. Neglecting the tensor and vector degrees of freedom, in Newtonian gauge, the line element takes the form
\eq{
	\label{eq:ds2}
	\dd{s}^2=-(1+2\Phi)\dd{t}^2+a^2(t)(1-2\Phi)\delta_{ij}\dd{x}^i\dd{x}^j\,,
}
where $\Phi (t, {\bf x})$ is the Newtonian potential. Since the anisotropic stress vanishes (to leading order in fluctuations) for an FLRW spacetime filled with a scalar field, the fluctuation
in the $00$-component of the metric is identical to the 
fluctuation in the spatial components. We split the field into a background component and small fluctuations, $\psi (t, {\bf x}) =\bpsi (t) +\dpsi (t, {\bf x} )$, where the splitting becomes unambiguous by requiring that the spatial average of $\dpsi (t, {\bf x})$ vanishes. The operator $\mathcal{D}$ defined in Eq.~\eqref{eq:D} up to linear order in $\Phi$ is
\eq{
\label{eq:DFRW}
\mathcal{D}=-\frac{3iH}{2}-\frac{\nabla^2}{2ma^2}+m\Phi+3iH\Phi+2i\dot{\Phi}\, .
} 
Hence, from Eqs.~(\ref{eq:psidotD}) and (\ref{eq:Tmn}), the equations governing the background evolution are
\eqa{
\label{eq:bckg1}
i\dot{\bpsi}+i\frac{3}{2}H\left(\bpsi-e^{2imt}\bpsi^*\right)&=0\\
\label{eq:bckg2}
3\Mpl^2H^2-m\bpsi^*\bpsi&=0\, .
}  
To linear order in spatially varying quantities, we find from the Klein-Gordon equation and the $0i$-component of Einstein's equations (after some algebra) the coupled equations
\eqa{
\label{eq:linpert1}
i\dot{\dpsi}+[\mathfrak{D}\dpsi-(m-2iH)\bpsi\Phi]
	+e^{2imt}[\mathfrak{D}^*\dpsi^*-(m+2iH)\bpsi^*\Phi]&=0 ,\\
\label{eq:linpert2}
\dot{\Phi}+H\Phi+\frac{i}{4\Mpl^2}\left(\bpsi\dpsi^*-\bpsi^*\dpsi+e^{-2imt}\bpsi\dpsi-e^{2imt}\bpsi^*\dpsi^*\right)&=0\, ,
}
in which we have defined a new operator
\eq{
\label{eq:Df}
\mathfrak{D}=\frac{3iH}{2}+\frac{\nabla^2}{2ma^2}-\frac{1}{2\Mpl^2}\left(e^{-imt}\bar{\psi}-e^{imt}\bar{\psi}^*\right)^2\,.
}
Note the difference between the new operator $\mathfrak{D}$ and the operator $\mathcal{D}$ in Eq.~\eqref{eq:D} for a general metric and its form for FLRW universe in Eq.~\eqref{eq:DFRW}. In deriving Eq.~\eqref{eq:linpert1} for $\dpsi$ we have used Eq.~\eqref{eq:linpert2} to eliminate the $\dot{\Phi}$ term appearing in Eq.~\eqref{eq:DFRW}. Further, note that $\Phi$ is a non-dynamical degree of freedom and can be eliminated by solving the $00$-component of Einstein's field equations,
\eq{
\label{eq:pois}
\frac{\nabla^2\Phi}{a^2}=\frac{m}{2\Mpl^2}(\bpsi\dpsi^*+\bpsi^*\dpsi)+3H\dot{\Phi}+
\frac{m}{4\Mpl^2}(e^{-imt}\bpsi+e^{imt}\bpsi^*)^2\Phi\, ,
}
which is the relativistic form of the Poisson equation. Here we have not explicitly eliminated the $\dot{\Phi}$ term from Eq.~\eqref{eq:linpert2} to avoid unnecessary complications, but we will do so in the calculations of the next section.  
Although it seems redundant to consider Eq.~\eqref{eq:linpert2} for $\Phi$, it would be much easier to obtain an effective description using Eq.~\eqref{eq:linpert2} rather than \eqref{eq:pois}. But we will keep track of Eq.~\eqref{eq:pois} and its effective form in the following sections since it helps us to interpret the results in the form of an effective imperfect fluid.   

Note again that the equations are so far exact, up to ${\cal O} (\Phi,\dpsi)$. However, in the nonrelativistic  (i.e. large mass) limit, we may neglect terms that are rapidly oscillating, since they average out to zero, as well as terms that are suppressed by factors of $(H/m)$. The above system of equations then reduces to the well-known Schr\"odinger-Friedmann and Schr\"odinger-Poisson equations for background and fluctuations, respectively
 \cite{Guth:2014hsa}:
\eqa{
\label{eq:leadbck}
	i\dot{\bpsi}+i\frac{3}{2}H\ \bpsi \simeq 0 \, , \qquad 
	3\Mpl^2H^2 =m\bpsi^*\bpsi \, ,
	\\
	i\dot{\dpsi}+\dfrac32 iH\delta \psi+\dfrac{\nabla^2}{2ma^2}\delta \psi -m\bpsi\Phi \simeq 0\, ,
\\
		\left(\frac{\nabla^2}{a^2} +\dfrac{3}{2}H^2\right)\Phi \simeq \frac{m}{2\Mpl^2}(\bpsi\dpsi^*+\bpsi^*\dpsi)\, ,
	}
where the second term on the left-hand side of the last equation is usually neglected, as these equations are usually considered for short (sub-horizon) scales. Note that the above equations are equivalent to a system of equations governing the evolution of a matter-dominated universe, showing that a nonrelativistic scalar field can play the role of cold dark matter (to leading order; see Sec.~\ref{sec:EFT} and Sec.~\ref{sec:fluid} for subleading deviations from this statement). 

We expect the above simplified equations to be valid to a good approximation. However we are interested in how subleading contributions change the dynamics of the approximated system. In particular, we want to take into account the backreaction of rapidly oscillating modes on the evolution of the slowly varying quantities. We therefore keep the subleading terms and later on will remove the fast oscillating modes in a systematic way to obtain an effective theory for slow modes. Note that, as discussed in the introduction, we are considering a system in which terms that vary on different characteristic time-scales couple to each other, which gives rise to the backreaction.
 
\section{The effective field theory in the nonrelativistic limit}
\label{sec:EFT}
In this section we outline the derivation and present the results of our effective field theory. We first introduce an appropriate mode decomposition (in time) to disentangle the part of the fields in which we are most interested (which varies slowly in time) from the parts that need to be systematically removed (which oscillate rapidly). We then discuss a procedure for removing the latter part, resulting in an effective theory for the former. We mainly follow the procedure outlined in Ref.~\cite{Namjoo:2017nia} (but see also Refs.~\cite{Mukaida:2016hwd,Braaten:2018lmj} for different, but equivalent, approaches). Here we consider the case in which the single field $\phi$ dominates the FLRW universe, and neglect any tree-level self-interactions. The effect of self-interaction shall be detailed in Appendix \ref{app:interaction} while the EFT for the case of a multicomponent universe is studied in Appendix \ref{app:multi-component}.  

\subsection{Smearing and mode expansion}\label{sec:mexp}

By applying the field redefinition introduced in the last section, in which we isolated the main time dependence of the field as an oscillatory factor (to leading order), we expect that apart from explicit factors like $\exp(2imt)$ appearing in the equations of motion, all remaining functions to be slowly varying in time. Let us denote the variables appearing in the equations of the last section collectively by $X(t)$, where we suppress any possible spatial dependence. By the above reasoning we expect $\dot{X}(t)\ll m X(t)$ with $m$ the mass of the scalar field. In other words, the spectrum of $X(t)$ in the frequency domain, $\hat X(\omega)$, would be localized around $\omega=0$. However, this cannot be the whole story because the explicit rapidly oscillating factors in the equation of motion, proportional to $\exp(2imt)$, will induce high-frequency components in the spectrum of $X(t)$. This happens through nonlinear multiplicative terms in Eqs.~\eqref{eq:bckg1}-\eqref{eq:pois}, which cause different frequencies to couple. Since the spectrum of the oscillatory factors like $e^{i \nu m t}$ (with $\nu$ an integer) would be a delta function at frequencies $\omega=\nu m$, we expect that at higher orders a localization happens around the frequencies $\omega=\nu m$. These subleading contributions would then backreact, again due to nonlinearities in the equations, on the slowly varying modes, whose spectra are localized around $\omega=0$. Our aim in this section is to study this backreaction to see how it affects the dynamics of the slow modes of different variables.

\begin{figure}
\begin{tikzpicture}[yscale=1.5]
\pgfmathdeclarefunction{sinc}{1}{%
        \pgfmathparse{abs(#1)<0.01 ? int(1) : int(0)}%
        \ifnum\pgfmathresult>0 \pgfmathparse{1}\else\pgfmathparse{sin(#1 r)/#1}\fi%
        }
\draw [thick , ->] (-4,0) -- (4,0);
\draw [thick , ->] (0,-0.5) -- (0,1.8);
\draw[domain=-3.8:3.8, samples=250,thick,blue] plot (\x,{sinc(3.5*\x)});

\node [above] at (0,1.8) {$W(t)$};
\node [right] at (4,0) {$t$};
\node at (0.3,1.15) {\small $\frac{m}{2\pi}$};
\node at (0.8,-0.18) {\small $\frac{2\pi}{m}$};

\end{tikzpicture}
\hspace{0.5 cm}
\begin{tikzpicture}[yscale=1.5]
\pgfmathdeclarefunction{rect}{1}{%
        \pgfmathparse{abs(#1)<1.5 ? 0.8 : int(0)}%
        }
\draw [thick ,->] (-3,0) -- (3,0);
\draw [thick ,->] (0,-0.5) -- (0,1.8);
\draw[domain=-3:3, samples=300,thick, blue] plot (\x,{rect(\x)});

\node [above] at (0,1.8) {$\hat{W}(\omega)$};
\node [right] at (3,0) {$\omega$};
\node at (1.5,-0.18) {\small $\frac{m}{2}$};
\node at (0.2,0.95) {\small $1$};
\end{tikzpicture}
\caption{The window function $W(t)$ (left) and its Fourier transform (right).}
\label{fig:window}
\end{figure}
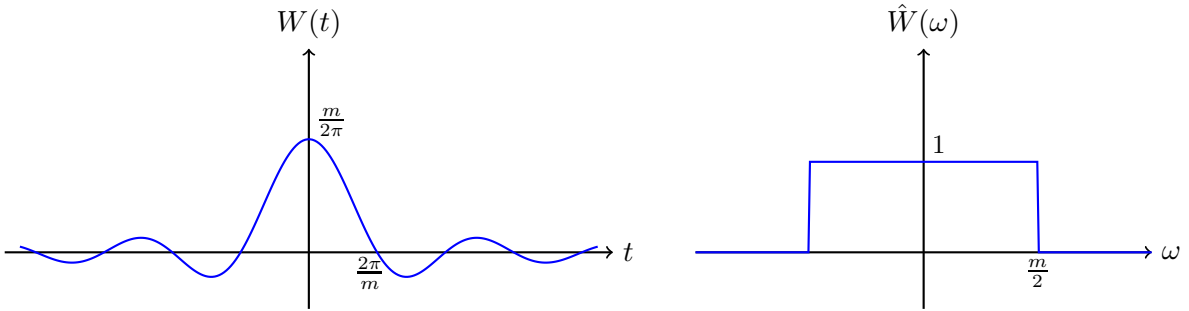

Since we are interested in the slow mode of functions of time (i.e. the portion of functions that are slow compared to oscillations with frequency $m$ or higher), as discussed in Sec.~\ref{sec:intro}, we can define a {\it smearing} operator acting on each function, which can be thought of as a time-average of the function, subject to a window function $W(t)$:
\eq{
\label{eq:avg}
	\ev{X(t)}=\int_{-\infty}^\infty \dd{t'}X(t')W(t-t')\, .
}  
This is the temporal counterpart of the spatial average considered for example in Ref.~\cite{Baumann:2010tm}. To adopt a suitable window function, note that in Fourier space, the right-hand side of Eq.~\eqref{eq:avg} takes the form ${\hat X}(\omega)\hat{W}(\omega)$.  As mentioned above, since the slow mode is localized around $\omega=0$ in the frequency domain, a natural choice for the window function would be a top-hat in Fourier space:
\eq{
{\hat W}(\omega)=
\begin{cases}
1\quad|\omega|<m/2\\
0\quad\text{otherwise}\,.
\end{cases}	
}   
Note that we have chosen the half-width of the square function to be $m/2$, which means that the window function removes the portion of the field that is spread away more than $m/2$ from each side of $\omega=0$. Choosing such a width is, first, consistent with requiring that the resulting smeared function has to be slowly varying compared to frequency $m$ and, second, leads to a convenient (and exact) mode expansion of $X(t)$, as we will explain below.
The window function in the time domain is then
\eq{
\label{eq:w}
W(t)=\int_{-\infty}^\infty \frac{\dd{\omega}}{2\pi}{\hat W}(\omega)e^{i\omega t}=\frac{\sin(mt/2)}{\pi t}\,.
}
Figure~\ref{fig:window} visualizes the window function in time and frequency domains. Note that the peculiar form of the window function in the time domain, with many zeros, is essential for capturing the low-frequency part (i.e. the slow mode) of the function. 

As mentioned earlier, in our nonlinear system, even if we start with purely slow functions of time, subdominant nonzero modes (which mainly vary with frequencies $\nu m$ with nonzero integer $\nu$) would develop. To capture these nonzero modes we can apply the same smearing operator and define the mode $\nu$ of the function by 
\eq{
	\label{eq:fnu_def}
	X_\nu(t)=\ev{X(t)e^{-i\nu mt}}=\int_{-m/2}^{m/2}\frac{\dd{\omega}}{2\pi}\hat{X}(\omega+\nu m)e^{i\omega t}\, ,
}
where in the last equality we explicitly used the adopted window function in the frequency domain. 
This result allows us to write any function of time by a mode expansion as follows: 
\eq{
	\label{eq:modexp}
	X(t)=\sum_{\nu=-\infty}^{\infty}X_\nu(t)e^{i\nu mt}\,.
}  
This is the key relation that will be used in the derivation of our EFT and from now on, our discussion will not explicitly rely on the smearing operator. Note that the specific choice of the window function makes the smearing operator a projection operator, i.e. $\ev{\ev{X(t)}}=\ev{X(t)}$. Note also that Eq.~\eqref{eq:modexp} is indeed exact, as a result of the appropriate choice of window function and its width; as can be checked by replacing the definition of $X_\nu$ from \eqref{eq:fnu_def} into \eqref{eq:modexp}. For later usage, note that the following relations hold for arbitrary functions of time $X(t)$ and $Y(t)$:
\eq{
	(XY)_\nu=\sum_{\alpha}X_\alpha Y_{\nu-\alpha}\,, \qquad 
	(X e^{i\mu m t})_\nu=X_{\nu-\mu}\, , \qquad (X^*)_\nu = X_{-\nu}^{*},
}
where the last equality should be understood as a relation between the modes of the complex conjugate of the field and the complex conjugate of the modes of the field. 
It is also easy to show that for a real function $R(t)$ we have $R_{-\nu}=R^*_\nu$. 

In the nonrelativistic limit, we will be interested in functions that are mainly localized around integer multiples of the characteristic frequency of the system $m$. In this regime, all coefficients $X_\nu(t)$ are slowly varying in time. Furthermore, in the same limit, the fields are mainly concentrated around $\omega=0$, i.e. the slow mode dominates over all other modes. This implies that the functions of time that we will be dealing with are slowly varying, so that a dimensionless parameter like  $\epsilon_t \sim |\dot{X}/mX|$ would always be small. Our EFT will then be an expansion in powers of such a  parameter, although other small parameters will come in as well, as discussed in the introduction and will be made explicit in the next section.

Now that we have a procedure to expand any function of time in terms of its different modes, we can derive differential equations for each mode. We are specially interested in the equation for the slow mode which, for later convenience, we denote by $X_s(t)\equiv X_{\nu=0}(t)$. The 
evolution of each mode involves all other modes due to nonlinearities in the equations. In the nonrelativistic limit, in which our time-dependent functions should be dominated by the slow mode, we require $|X_s| \gg |X_\nu|$ for $\nu \neq0$. In this regime, we can solve equations for nonzero modes perturbatively in the small quantities discussed in the Introduction. To achieve this, we further decompose each nonzero mode by a perturbative expansion
\eq{
\label{eq:pertexp}
X_\nu=\sum_{n=1}^{\infty}X_\nu^{(n)}= X_\nu^{(1)}+X_\nu^{(2)}+\dots\,, \qquad {\text{for}} \, \nu \neq 0,
} 
where for each $n$ we have $X_\nu^{(n)}=\order{\epsilon^n}$ and $\epsilon$ is a small parameter, to be determined by the equations governing the dynamics of the system, as in Eqs.~\eqref{eq:eps} and \eqref{eq:ept}. We always work up to the same order in all small parameters so that we can schematically show the orders in a perturbative expansion by just one parameter $\epsilon$ as above. After obtaining the solutions of the nonzero modes (in terms of the slow modes) order by order in perturbation theory, we plug the solutions back into the equations for the slow mode. This procedure leads to an effective equation for the slow mode. The tools that we have developed here (mostly implicitly) will allow us to obtain EFTs for the background evolution as well as for the small fluctuations, as we shall show explicitly below.    

\subsection{Effective field theory for the background}\label{sec:eftbkg}
In this section we apply the procedure we outlined in the last section to the equations of motion for the spatially homogeneous scalar field in an FLRW background, Eqs.~\eqref{eq:bckg1} and \eqref{eq:bckg2}. For convenience, we introduce the rescaled, dimensionless quantities 
\eq{
\label{eq:rescale}
\cpsi (t) \equiv\frac{\bar{\psi} (t) }{\sqrt{m}\Mpl}\,,\qquad\cH (t) \equiv\frac{H (t)}{m}\,,
}
which we use in the course of calculations. We also rescale the time $\tilde{t}=mt$. Application of the mode expansion introduced in the previous section to Eqs.~\eqref{eq:bckg1} and \eqref{eq:bckg2}, with rescaled variables, results in the following equations for each mode $\nu$:
\eqa{
	\label{eq:psinu}
	\cpsi'_\nu+
	i\nu \cpsi_\nu+\frac{3}{2}\cH_\alpha(\cpsi_{\nu-\alpha}-\cpsi^*_{2+\alpha-\nu})&=0\\
	\label{eq:Hnu}
	3\cH_\alpha \cH_{\nu-\alpha}-\cpsi_\alpha \cpsi^*_{\alpha-\nu}&=0\,,
}
where the summation over repeated indices is understood and  the prime denotes the derivative with respect to $\tilde{t}$. We are, in particular, interested in the equations for the slow mode quantities $\bar{\psi}_s=\ev{\bar{\psi}}$ and $H_s=\ev{H}$, with the corresponding equations after rescaling the variables:
\eq{
\label{eq:psis0}
\cpsi'_s+\frac{3}{2}\bigg(\cH_s\cpsi_s+\sum_{\alpha\neq0}\cH_\alpha\cpsi_\alpha\bigg)-\frac{3}{2}\bigg(\cH_s\cpsi_2^*+\cH_{-2}\cpsi_s^*+\sum_{\alpha\neq0,-2}\cH_\alpha\cpsi^*_{\alpha+2}\bigg)=0\,,
}
and
\eq{
\label{eq:Hs0}
3\bigg(\cH_s^2+\sum_{\alpha\neq0}|\cH_\alpha|^2\bigg)-\bigg(|\cpsi_s|^2+\sum_{\alpha\neq0}|\cpsi_\alpha|^2\bigg)=0\,,
}
which are obtained simply by setting $\nu=0$ in Eqs.~\eqref{eq:psinu} and \eqref{eq:Hnu}. We see clearly from the last set of equations that the slow  mode of each function is coupled to all nonzero modes due to nonlinearities of the system. We can solve the equations for the nonzero modes perturbatively in small parameters. As discussed in the introduction, we identify a new small parameter, in addition to $\epsilon_t$, from the Hubble scale
\eq{
\epsilon_{\scriptscriptstyle H}\sim\frac{H}{m} =\tilde H\,.
}
It is evident that such a quantity has to be small for a nonrelativistic field rapidly oscillating around the minimum of its potential. 
Thus, at the background level, we have two small parameters in the problem, which we collectively denote by $\epsilon=\{\epsilon_t,\epsilon_{\scriptscriptstyle H}\}$. In this section we will keep terms up to third order in $\epsilon$ but our formalism is general and can be continued to higher orders. Note that the rescaled variables introduced in Eq.~\eqref{eq:rescale} are first order in $\epsilon$, i.e. $\cpsi_s=\order{\epsilon}$ and $\cH_s=\order{\epsilon}$, which makes the power counting in the derivation of EFT fairly easy. We also have $\cpsi_s'\sim\epsilon\cpsi_s=\order{\epsilon^2}$. It is important to notice that working up to order $n$ in solving for nonzero modes would result in an EFT for the slow modes up to $\order{\epsilon^{n-1}}$ corrections once the corrections are compared to leading order terms. This is because even the zeroth-order equations for the slow modes are suppressed by a factor of $\order{\epsilon}$ (due to, for example, the time derivative operator or the Hubble parameter). The same statement will be valid for fluctuations as well.

To proceed systematically, we apply the perturbative expansion of \eqref{eq:pertexp} to the nonzero modes. Note that we do not apply the perturbative expansion to the slow modes since we are interested in the effective equations for those quantities, rather than their corresponding solutions. From \eqref{eq:psinu} and for $\nu\neq0$ we have 
\eq{
\label{eq:psinun}
\spl{
\cpsi_\nu^{(n)}=&-\frac{\cpsi_\nu^{(n-1)'}}{i\nu}-\frac{3}{2i\nu }\bigg[\cH_s\cpsi_\nu^{(n-1)}+\cH_\nu^{(n-1)}\cpsi_s+\sum_{\ell=1}^{n-2}\sum_{\alpha\neq0,\nu}\cH_{\alpha}^{(\ell)}\cpsi_{\nu-\alpha}^{(n-\ell-1)}\bigg]\\
&+\frac{3}{2i\nu}\bigg[\cH_s\cpsi_s^*\delta_{\nu,2}\delta_{n,2}+(1-\delta_{\nu,2})\big(\cH_s\cpsi_{2-\nu}^{(n-1)}{}^*+\cH_{\nu-2}^{(n-1)}\cpsi_s^*\big) +\sum_{\ell=1}^{n-2}\sum_{\alpha\neq0,\nu-2}\cH_{\alpha}^{(\ell)}\cpsi_{2+\alpha-\nu}^{(n-\ell-1){}^*}\bigg]\,,
}
}
where $n\geq1$. Similarly From Eq.~\eqref{eq:Hnu} for $\nu\neq0$ we get
\eq{
\label{eq:Hnun}
\cH_\nu^{(n)}=\frac{1}{6\cH_s}\bigg[\cpsi_s\cpsi^{(n)}_{-\nu}{}^*+\cpsi_s^*\cpsi^{(n)}_{\nu}+\sum_{\ell=1}^{n}\sum_{\alpha\neq0,\nu}\cpsi_\alpha^{(\ell)}\cpsi_{\alpha-\nu}^{(n+1-\ell)}{}^*\bigg]-\frac{1}{2\cH_s}\sum_{\ell=1}^{n}\sum_{\alpha\neq0,\nu}\cH_\alpha^{(\ell)}\cH_{\nu-\alpha}^{(n+1-\ell)}\,.
} 
Note that since the rescaled variables $\cpsi$ and $\cH$ are first order in $\epsilon$ we expect the nonzero modes $\cpsi_\nu$ and $\cH_\nu$ to start from the second order. We will see below that it is in fact the case. 
By using Eqs.~\eqref{eq:Hnun} and \eqref{eq:psinun} we see that 
\eq{
\cpsi_\nu^{(1)}=\cH_\nu^{(1)}=0\,,
}
and for the leading nonzero modes we find
\begin{eBox}
\eq{
\label{eq:psinu1}
\cpsi_\nu^{(2)}=
\frac{3}{4i}\cH_s\cpsi_s^* \, \delta_{\nu,2}\, , \qquad 
\cH_\nu^{(2)}=\frac{1}{8i}\cpsi_s^*{}^2 \, \delta_{\nu,2}-\frac{1}{8i}\cpsi_s^2 \, \delta_{\nu,-2}\,.
}
\end{eBox}
In obtaining this result there was no need to solve a differential equation, since the time derivative term is  subleading, contributing to higher orders. The relation between nonzero modes and the zero modes is one of the main results of our paper.
%
%
We can easily proceed and compute higher-order terms in $\epsilon$. After some algebra we find that for $n=3$, 
\eq{
\spl{
\cpsi_\nu^{(3)}=&\left(\frac{3}{8}(\cH_s\cpsi_s^*)'+\frac{9}{16}\cH_s^2\cpsi_s^*+\frac{3}{32}|\cpsi_s|^2\cpsi_s^*\right) \, \delta_{\nu,2}+\frac{3}{32}\cpsi_s^3 \, \delta_{\nu,-2}-\frac{3}{64}\cpsi_s^*{}^3 \, \delta_{\nu,4}\,,
}
}
and similarly for the Hubble parameter,
\eq{
\spl{
\cH_\nu^{(3)}=&\frac{1}{6\cH_s}\left(\frac{3}{8}(\cH_s\cpsi_s^*)'\cpsi_s^*+\frac{9}{16}\cH_s^2\cpsi_s^*{}^2+\frac{3}{32}|\cpsi_s|^4+\frac{3}{32}\cpsi_s^4\right) \, \delta_{\nu,2}\\
&+\frac{1}{6\cH_s}\left(\frac{3}{8}(\cH_s\cpsi_s)'\cpsi_s+\frac{9}{16}\cH_s^2\cpsi_s{}^2+\frac{3}{32}|\cpsi_s|^4+\frac{3}{32}\cpsi_s^*{}^4\right) \, \delta_{-\nu,2}\,.
}
}
 Since in this section we will only consider corrections up to order $n=3$ for the background equations, we can use the leading-order equations \eqref{eq:leadbck} for the slow modes  to simplify the above relations to the same order, which yields
\begin{eBox}
\eq{
\label{eq:psinu2}
\cpsi_\nu^{(3)}=-\frac{3}{32}|\cpsi_s|^2\cpsi_s^* \, \delta_{\nu,2}+\frac{3}{32}\cpsi_s^3 \, \delta_{\nu,-2}-\frac{3}{64}\cpsi_s^*{}^3 \, \delta_{\nu,4}\,,\qquad\cH_\nu^{(3)}=0\,.
}
\end{eBox}
Now that we have obtained the solutions for nonzero modes, we may substitute into the equations of motion for the slow modes, Eqs.~\eqref{eq:psis0} and \eqref{eq:Hs0}. Considering terms up to order $n=3$ we have
\begin{equation}
\label{eq:psisorders}
\begin{split}
&\cpsi_s'
+\frac{3}{2}\cH_s\cpsi_s
\\
&-\frac{3}{2}\left(\cH_s\cpsi_2^{(2)}{}^*+\cH_{-2}^{(2)}\cpsi_s^*\right)\\
&-\frac{3}{2}\bigg(\cH_s\cpsi_2^{(3)}{}^*+\cH_{-2}^{(3)}\cpsi_s^*-\cH_2^{(2)}\cpsi_2^{(2)}\bigg)+\order{\epsilon^5}=0\,,
\end{split}
\end{equation}
and
\begin{equation}
\begin{split}
&3\cH_s^2-|\cpsi_s|^2\\
&+3\left(\cH_{2}^{(2)}\right)^2+3\left(\cH_{-2}^{(2)}\right)^2-\vert \cpsi_2^{(2)}\vert^2+\order{\epsilon^5}=0\,,
\end{split}
\end{equation}
where we have written the terms that are at the same order in the same line and we have excluded the terms that contain nonzero modes at order $n=1$ as they vanish identically.  After substitution of nonzero modes we finally obtain the effective equations of motion for the slow modes. For convenience we write the equations in this last step using the original, dimensionful variables:
\begin{eBox}
\eqa{
\label{eq:EFTbck1}
&i\dot{\bar{\psi}}_s+i\frac{3}{2}H_s\bar{\psi}_s+\frac{9}{16}\frac{|\bar{\psi}_s|^2}{\Mpl^2} \bar{\psi}_s+i\frac{9}{32}\frac{H_s}{m}\frac{|\bar{ \psi}_s|^2}{\Mpl^2}\bar{\psi}_s+\order{\epsilon^3}H_s\bar{\psi}_s=0 \, ,\\
\label{eq:EFTbck2}
&3\Mpl^2H_s^2=m|\bar{\psi}_s|^2+\frac{3}{32\Mpl^2}|\bar{\psi}_s|^4+\order{\epsilon^4}m|\bar{\psi}_s|^2\,.
} 
\end{eBox}
This is one of our main results in this paper. Note that, interestingly, the backreaction of the nonzero modes on the slow mode effectively induces a kind of self-interaction for the slow mode and a new contribution to the energy density, which sources the Hubble parameter. (We will consider characteristics of the induced self-interaction below, after deriving our EFT for the fluctuations.) We expect these nontrivial corrections to the Schr\"odinger and Friedmann equations to help improve the validity of the simplified nonrelativistic equations when compared to the exact equations. In Sec.~\ref{sec:num}, where we present the numerical solutions, we will confirm  that this is indeed the case, and illustrate that the solutions of our EFT follow the exact solutions closely. We can also provide further insight into the above results by describing the system as a fluid in an expanding background, which will be done in Sec.~\ref{sec:fluid}. 

\subsubsection*{The scale factor}
Since in a flat FLRW universe, the background equations do not depend on the scale factor explicitly, we did not need to solve for it. However, since the scale factor would show up in the equations governing small fluctuations, we would need to have a perturbative expansion of it as well. A simple way to achieve this is to use its differential equation $\dot{a}=aH$, which, after the mode expansion and using the rescaled variables, results in   
\begin{equation}
a'_\nu+i\nu a_\nu=a_\alpha \cH_{\nu-\alpha}\,.
\end{equation}
As before, we perform a perurbative expansion for nonzero modes, and it is easy to see that $a_\nu^{(1)}=0$ for all $\nu$. For $n>1$ we have
\begin{equation}
a_\nu^{(n)}=-\frac{a_\nu^{(n-1)}{'}}{i\nu}+\frac{a_s\cH_\nu^{(n)}}{i\nu}+\frac{a_\nu^{(n-1)}\cH_s}{i\nu}+\sum_{\ell=1}^{n-1}\sum_{\alpha\neq\{0,\nu\}}\frac{a_\alpha^{(\ell)}\cH_{\nu-\alpha}^{(n-\ell)}}{i\nu}\,.
\end{equation}
Specifically, for $n=2$ the we have
\begin{eBox}
\begin{equation}
a_\nu^{(2)}=\frac{1}{2i}\cH_\nu^{(2)}a_s=\left(-\frac{1}{16}\cpsi_s^*{}^2 \, \delta_{\nu,2}-\frac{1}{16}\cpsi_s^2 \, \delta_{\nu,-2}\right)a_s\,.
\end{equation}
\end{eBox}
Since, for the fluctuations, we work up to $n=2$ in perturbative expansion, we only need the scale factor up to this order. Therefore, we neglect higher-order corrections to the scale factor.  
Finally, the effective equation for the slow-mode scale factor $a_s=\ev{a}$ to this order becomes
\begin{equation}
\dot{a}_s=a_s H_s+\order{\epsilon^2}a_sH_s\, ,
\end{equation}
that is, to this order, the slow mode of the scale factor does not receive any correction from nonzero modes. Before concluding this section, we define variables for the inverse of the scale factor and its square for later use:  
\eq{
\label{eq:r}
q(t)\equiv\frac{1}{a(t)}\,,\qquad r(t)\equiv\frac{1}{a(t)^2}\,.
}
Then it is straightforward to show that $q_\nu^{(1)}=r_\nu^{(1)}=0$ and for $n=2$ we have $q_\nu^{(2)}=-a_\nu^{(2)}/a_s^2$ and $r_\nu^{(2)}=-2a_\nu^{(2)}/a_s^3$.

\subsection{Effective field theory for fluctuations}\label{sec:eftpert}
We now proceed to find the effective equations of motion for the slow modes of the linear fluctuations around FLRW background. The procedure is similar to the last section. We start by applying the mode expansion to Eqs.~\eqref{eq:linpert1} and \eqref{eq:linpert2}. Note that we need to use background quantities and their mode expansion as well. We again use the rescaled variables defined in Eq.~\eqref{eq:rescale} as well as
\eq{
\cdpsi (t, {\bf x} )\equiv\frac{\dpsi (t, {\bf x}) }{\sqrt{m}\Mpl}\,.
}
Furthermore, to make all the expressions dimensionless (which makes the power counting more straightforward) we rescale the spatial coordinates by $\tilde{x}=mx$ and use the notation $\cnabla^2$ for the corresponding comoving spatial Laplacian operator. Our aim is to compute the effective equations of motion for $\dpsi_s=\ev{\dpsi}$ and $\Phi_s=\ev{\Phi}$. The relevant equations for general mode $\nu$ from Eqs.~\eqref{eq:linpert1} and \eqref{eq:linpert2} in rescaled variables read
\eqa{
\label{eq:dpsinu}
&i\cdpsi'_\nu-\nu\cdpsi_\nu+(\tilde{\mathfrak{D}}_\alpha\cdpsi_{\nu-\alpha}+\tilde{\mathfrak{D}}^*_{-\alpha}\cdpsi^*_{2+\alpha-\nu}) \nonumber \\ 
&\quad\quad\quad  -\Phi_{\alpha}(\cpsi_{\nu-\alpha}+\cpsi^*_{2+\alpha-\nu}-2i\cH_\beta(\cpsi_{\nu-\alpha-\beta}-\cpsi^*_{2+\alpha+\beta-\nu}))=0\,,\\
\label{eq:Phinu}
&\Phi'_\nu+i\nu \Phi_\nu+\cH_\alpha\Phi_{\nu-\alpha}-\frac{1}{4i}(\cpsi_\alpha\cdpsi_{\alpha-\nu}^*-\cpsi_{-\alpha}^*\cdpsi_{\nu-\alpha}+\cpsi_\alpha\cdpsi_{\nu+2-\alpha}-\cpsi^*_{-\alpha}\cdpsi^*_{2+\alpha-\nu})=0\,,
}
where the rescaled operator is
\eq{
\tilde{\mathfrak{D}}_\nu=\frac{1}{2}\Big(3i\cH_\nu-r_\nu\cnabla^2+2\cpsi^*_{-\beta}\cpsi_{\nu-\beta}-\cpsi_\beta\cpsi_{\nu+2-\beta}-\cpsi^*_{-\beta}\cpsi_{2+\beta-\nu}\Big)\,,
}
and we have used the notation defined in Eq.~\eqref{eq:r} for the mode expansion of $1/a^2$.
The Poisson equation, Eq.~\eqref{eq:pois}, leads to
\begin{equation}
\spl{
r_\alpha\tilde{\nabla}^2\Phi_{\nu-\alpha}=&\frac{1}{2}(\cpsi^*_{-\alpha}\cdpsi_{\nu-\alpha}+\cpsi_{\alpha}\cdpsi^*_{\alpha-\nu})+3\cH_{\nu-\alpha}(\Phi_{\alpha}'+i\alpha\Phi_{\alpha})\\
&+\frac{1}{4}(\cpsi_\alpha\cpsi_{\nu+2-\alpha-\beta}+\cpsi^*_{-\alpha}\cpsi^*_{2+\alpha+\beta-\nu}+2\cpsi^*_{-\alpha}\cpsi_{\nu-\alpha-\beta})\Phi_{\beta}\,.
}
\end{equation}
Note that we can eliminate the $\Phi'_\alpha$ term by using Eq.~\eqref{eq:Phinu}.

Once again we have an infinite set of coupled equations and to proceed we have to identify small parameters in the problem. Besides the previously introduced parameters $\epsilon_{\scriptscriptstyle H}$ and $\epsilon_t$, there is another one due to the fact that we are now dealing with field fluctuations with nonzero momenta. In the nonrelativistic limit we expect $\epsilon_k \sim (k^2/m^2a^2)$ to be small, where $k$ is the comoving momentum of the nonrelativistic field. Furthermore, in order to make the power counting in the perturbative expansion tractable, we take into account the fact that the rescaled fluctuations are small  (compared to the background quantities), which we quantify by yet another small parameter $\epsilon_g$, so that we have $\cdpsi=\order{\epsilon_g}$. It is important to note that since we are working only to linear order in spatially varying quantities, the new parameter $\epsilon_g$ becomes irrelevant. Nevertheless, introducing this parameter is helpful for the perturbative expansion of nonzero modes, when we must evaluate the size of each term. Therefore, our EFT for the field fluctuations would be an expansion in four small parameters, $\epsilon=\{\epsilon_t,\epsilon_{\scriptscriptstyle H},\epsilon_k,\epsilon_g\}$. Note that we have not made any assumption about the possible hierarchy between these parameters. In particular, fluctuations can be both at sub-horizon and super-horizon scales without affecting the following analysis, as long as the Hubble parameter and the physical wavenumber are small compared to the mass of the field. Note that in terms of rescaled variables, we have a useful order of magnitude relation $\cdpsi_s\sim\Phi_s=\order{\epsilon}$.   

Similar to what has been done in the previous section, by using the perturbative expansion in Eq.~\eqref{eq:pertexp}, we can solve for the relevant nonzero modes and find the effective equations of motion for the slow modes. Again, similar to the background case, we can deduce that at order $n=1$ we have
\eq{
\cdpsi_\nu^{(1)}=\Phi_\nu^{(1)}=0\,,
}
and the leading nonzero modes corresponding to $n=2$ can be obtained by using Eqs.~\eqref{eq:dpsinu} and \eqref{eq:Phinu}
\begin{eBox}
\eq{
\label{eq:Phinu2}
\cdpsi_{\nu}^{(2)}=\left(\frac{3}{4i}\cH_s\cdpsi_s^*+\frac{\cnabla^2\cdpsi_s^*}{4a_s^2}-\frac{1}{2}\cpsi_s^*\Phi_s\right)\delta_{\nu,2}\,,\quad
\Phi_\nu^{(2)}=\frac{1}{8}\bigg(\cpsi_s^*\cdpsi_s^*\delta_{\nu,2}+\cpsi_s\cdpsi_s\delta_{\nu,-2}\bigg)\,.
}
\end{eBox}
We will present the equations for higher orders ($n>2$) in Appendix~\ref{app:linhigher} and limit ourselves to order $n=2$ (and hence up to $\order{\epsilon}$ corrections to the equations for the slow-modes) in this section. As we will see, even at this order we find nontrivial terms in the effective equations of motion.
%
Having found the solutions for the leading-order nonzero modes, the effective equations governing the slow mode fluctuations $\dpsi_s$ and $\Phi_s$ can be obtained by setting $\nu=0$ in Eqs.~\eqref{eq:dpsinu} and \eqref{eq:Phinu} and replacing the nonzero modes of the background and fluctuation variables by their corresponding solutions. Returning to the original variables at the last step, this procedure results in
\begin{eBox}
\eq{
\spl{
	\label{eq:eftsch}
	&i\dot{\dpsi_s}+i\frac{3}{2}H_s\dpsi_s+\frac{\nabla^2\dpsi_s}{2ma_s^2}-m\bar{\psi}_s\Phi_s\\
	&+\frac{9}{8}\frac{|\bar{ \psi}_s|^2}{\Mpl^2} \dpsi_s-\frac{7}{16}\frac{\bar{\psi}_s^2}{\Mpl^2}\dpsi_s^*+\frac{\nabla^4\dpsi_s}{8m^3a_s^4}+2iH_s \bar{\psi}_s \Phi_s+\order{\epsilon^2}H_s\delta\psi_s=0\,,
}
}
\end{eBox}
as the effective Schr\"odinger equation, and 
\begin{eBox}
\eq{
\spl{
	\label{eq:eftphidot}
	&\dot{\Phi}_s+H_s\Phi_s+\frac{i}{4\Mpl^2}(\bar{\psi}_s \dpsi_s^*-\bar{\psi}_s^*\dpsi_s)\\
	&+\frac{3H_s}{8m\Mpl^2}(\bar{\psi}_s\dpsi_s^* + \bar{\psi}_s^*\dpsi_s)+\frac{i}{8m\Mpl^2}\frac{\nabla^2}{2ma_s^2}(\bar{\psi}_s \dpsi_s^*- \bar{\psi}_s^*\dpsi_s)+\order{\epsilon^2}H_s\Phi_s=0\,,
}
}
\end{eBox}
as the equation for the slow-mode Newtonian potential $\Phi_s$. We also have the effective Poisson equation,
\begin{eBox}
\eq{
\spl{
\label{eq:eftpoisson}
\frac{\nabla^2\Phi_s}{a_s^2}=&\frac{m}{2\Mpl^2}(\bar{\psi}^*_s \dpsi_s+ \bar{\psi}_s\dpsi^*_s)\\
&+\frac{3i}{4\Mpl^2}H_s(\bar{\psi}^*_s \dpsi_s- \bar{\psi}_s\dpsi^*_s)-\frac{3}{2}H_s^2\Phi_s+\order{\epsilon^2}H_s^2\frac{\dpsi_s}{\bar{\psi}_s}\,.
}
}
\end{eBox}
Once again we see that the nonzero modes induce nontrivial corrections to all equations for the slow modes. We stress that all these corrections have been mostly neglected in the literature to date. In Sec.~\ref{sec:fluid}, we provide an alternative description of the system in terms of an imperfect fluid, to extract useful effective quantities for the system, such as the sound speed and viscosity. Also, in Sec.~\ref{sec:num}, we confirm the validity of our EFT by numerical analysis. In Appendix~\ref{app:multi-component} we study the case of a multicomponent universe and in Appendix~\ref{app:interaction} we take into account the self-interaction of the scalar field, which has been neglected so far. 

In both Eq.~\eqref{eq:EFTbck1} for $\bar{\psi}_s$ and Eq.~\eqref{eq:eftsch} for $\delta \psi_s$, we find nonlinear terms in the effective equations of motion that take the form of self-interaction terms, even though we began with a free scalar field. To further characterize the induced self-interaction, we may consider Eqs.~\eqref{eq:eftbckslf} and \eqref{eq:fluctslambda} for the evolution of the slow modes $\bar{\psi}_s$ and $\delta \psi_s$ in the presence of a tree-level self-interaction of the form $V (\phi) = \lambda \phi^4 / (4!)$ for the original (relativistic) scalar field $\phi$. If we consider $\psi_s (t, {\bf x}) = \bar{\psi}_s (t) + \delta \psi_s (t, {\bf x})$ and work to first order in fluctuations, we may combine Eqs.~\eqref{eq:eftbckslf} and \eqref{eq:fluctslambda} to write
\begin{equation}
\begin{split}
i \dot{\psi}_s &+ \frac{3i}{2} H_s \psi_s + \frac{\nabla^2 \psi_s}{2m a_s^2}  - \frac{\lambda}{8 m^2} \vert \psi_s \vert^2 \psi_s + \frac{ 9}{16 \Mpl^2} \vert \psi_s \vert^2 \psi_s   \\
&\quad + \frac{\nabla^4 \psi_s}{8 m^3 a_s^4}  - m \bar{\psi}_s \left[ 1 - 2i \left( \frac{ H_s}{m} \right) \right] \Phi_s + {\cal O} (\epsilon^3 ) m \psi_s = 0.
\end{split}
\label{eq:eompsicombined}
\end{equation}
(In the limit of a rigid spacetime, with $a_s (t) \rightarrow 1$, $(m / \Mpl) \rightarrow 0$, and $\vert \Psi_s (t, {\bf x}) \vert \rightarrow 0$, Eq.~\eqref{eq:eompsicombined} matches the equation of motion found in Ref.~\cite{Namjoo:2017nia} for the slow mode $\psi_s (t, {\bf x})$ in Minkowski spacetime, to the appropriate order.) The form of Eq.~\eqref{eq:eompsicombined} suggests that the {\it induced self-interaction strength}, even in the absence of a tree-level self-coupling, takes the form
\begin{equation}
    \lambda_{\rm eff} = - \frac{9}{2} \left( \frac{m}{\Mpl} \right)^2 + {\cal O} (\epsilon^2) .
    \label{eq:lambdaeff}
\end{equation}
We see immediately that the induced self-interaction is attractive $(\lambda_{\rm eff} < 0$) and that it arises gravitationally (being proportional to $G \sim 1/\Mpl^2$).\footnote{The right-hand side of the effective Friedmann equation in Eq.~\eqref{eq:EFTbck2} includes a contribution to $\rho_{\rm eff}$ proportional to $\lambda_{\rm eff} \vert \bar{\psi}_s \vert^4$, scaling with $\lambda_{\rm eff}$ and $\bar{\psi}_s$ in the way one would naively expect from the form of the induced self-coupling, though with a different sign and overall coefficient. The term proportional to $\vert \bar{\psi}_s \vert^4$ in Eq.~\eqref{eq:EFTbck2} arises from backreaction of rapid oscillations on $H_2^{(2)}$ and $H_{-2}^{(2)}$. How best to incorporate global gravitational effects arising from the gravitationally induced self-interactions remains the subject of further study.} 

The emergence of a gravitationally induced self-coupling could be of interest in the context of self-interacting dark matter \cite{Bullock:2017xww,Fan:2016rda,Tulin:2017ara}, though estimating the magnitude of the effects remains subtle. To address various astrophysical observations, models of self-interacting dark matter typically introduce interactions that yield $\sigma / m \lesssim 1 \, {\rm cm}^2 / g = 4.6 \times 10^3 \, {\rm GeV}^{-3}$, where $\sigma$ is the total scattering cross section. Given the form of $\lambda_{\rm eff}$ in Eq. (3.41), the induced self-interactions we have identified appear, at least naively, to be much too weak to account for such interactions. However, a direct comparison between the cross section deduced from our EFT and the cross section required in self-interacting dark matter models is complicated for at least two reasons. First, our EFT calculation is only valid up to linear order in spatially varying fluctuations, whereas the self-interactions within dark matter halos require a fully nonlinear analysis. Second, the very high occupation numbers of particles in ultra-light dark matter models implies that classical field equations yield a more reliable description of the system \cite{Deng:2018jjz} than naive scattering amplitudes calculated within quantum field theory. We leave a more systematic study of this interesting topic for future work. As we will see in the next section, however, despite the small magnitude of the gravitationally induced self-coupling, the backreaction effects captured in our effective description lead to interesting -- and in principle observable -- phenomenological features, including nontrivial pressure, sound speed, and bulk viscosity.

As a final --- but important --- remark, notice that a by-product of our EFT is the ability to construct the full solution: Once the slow modes are found by (perhaps numerically) solving the effective equations, we are able to find the full solution, including its oscillatory behavior. For example, the scalar field $\psi$ can be constructed via $\psi = \sum_\nu \psi_\nu\, e^{i\nu mt}$, where $\psi_\nu$ for $\nu\neq 0$ are obtained order by order by our perturbative prescription and are given once the solution for $\psi_s$ is obtained from the EFT.  In Sec.~\ref{sec:num} we show that this procedure leads to results that match to the exact solution with very good accuracy (and one can obtain even more accurate results by going to higher orders in the perturbation theory). In situations where the oscillatory behavior is of interest, this procedure is expected to be much more efficient compared to solving the exact equations. This is because the exact equations of motion, involving rapidly oscillating factors, are generically expected to suffer from stiffness and instabilities. On the other hand, the EFT equations are well behaved and are expected to be solved easily by standard numerical algorithms. Such a theoretical framework would be appropriate for a variety of situations, relevant to cosmology and astrophysics. Examples of such situations are the change in the orbits of planets and stars in the dark matter halo \cite{Boskovic:2018rub} or the resonances in binary pulsars \cite{Blas:2019hxz} as results of the  oscillations of dark matter.

\section{Effective fluid description} 
\label{sec:fluid}
Having derived an EFT for the system, it is instructive to find a more intuitive way to interpret the results. In this section, we obtain an equivalent description of the system in terms of an imperfect fluid and argue that the higher-order terms in our EFT can be understood as new contributions to the variables describing the fluid. 

To start, we may interpret the equations \eqref{eq:EFTbck1} and \eqref{eq:EFTbck2} as a universe (effectively) expanding with rate $H_s$ and filled with some effective fluid. As a result, we may consider the right-hand side of the Friedmann equation, Eq.~\eqref{eq:EFTbck2}, as an effective energy density of the fluid: 
\begin{eBox}
\eq{
	\rho_{\text{eff}} = m \vert \bar{\psi}_s \vert^2 +\dfrac{3}{32\Mpl^2}\vert \bar{\psi}_s \vert^4\,.
	\label{eq:rho_eff}
}
\end{eBox}
Using the Schr\"odinger equation, Eq.~\eqref{eq:EFTbck1}, we can then derive a continuity equation for the fluid in an FLRW background,
\eq{
	\dot \rho_{\text{eff}} +3H_s(\rho_{\text{eff}}+p_{\text{eff}})=0\,,
}
where the effective pressure turns out to be
\begin{eBox}
\begin{equation}
p_{\text{eff}}=\frac{9}{32 \Mpl^2}|\bar{\psi}_s|^4\,.
\label{peff}
\end{equation}
\end{eBox}
Notice that the effective pressure is of order $\sim\order{\epsilon^2}\rho_{\text{eff}}$, so to $\order{\epsilon}\rho_{\text{eff}}$ the effective fluid behaves like pressureless matter, consistent with cold dark matter.\footnote{ The effective pressure in Eq.~(\ref{peff}) should not be confused with the smeared pressure of the scalar field appearing in the energy-momentum tensor at the level of the background, $\ev{p_\phi}= \langle \frac{1}{2} \dot \phi^2 -\frac{1}{2}m^2 \phi^2 \rangle$, as is done for example in Ref.~\cite{Blas:2019hxz}. The reason is that in the former, we absorbed another contribution from the left-hand side of the Friedmann equation, which appears as a result of integrating out the nonzero modes of the Hubble parameter, such as $H_2$ and $H_{-2}$. } Nonetheless, the rapidly oscillating modes induce a small effective pressure, which yields an effective equation of state $w_{\rm eff} = p_{\rm eff} / \rho_{\rm eff}$ of the form
\begin{equation}
    w_{\rm eff} = \frac{ 27}{32} \left( \frac{ H_s }{m} \right)^2 + {\cal O} (\epsilon^3) ,
\end{equation}
upon using Eq.~(\ref{eq:EFTbck2}) for $H_s$. It is evident that the effective pressure is a purely gravitational effect which induces a sort of interaction in the fluid, even though the original theory involves only a free scalar field.  


We aim to make a similar analogy between field fluctuations in our EFT and fluctuations of the fluid. Once we include spatially varying field fluctuations, our corresponding fluid description will feature an imperfect fluid. For such a description, we must incorporate the bulk viscosity, parameterized by the coefficient $\zeta$.\footnote{In principle, we expect other variables in the imperfect fluid, such as the shear viscosity and the effect of heat transfer, to appear as well. This is because such contributions are consistent with --- and hence allowed by --- the symmetries of the problem. However, as we will see, these additional variables are not required to fully describe the low-energy system under study to working order, although they may show up at higher orders, neglected in this paper.}
In Appendix~\ref{app:vis} we present some equations governing an imperfect fluid with bulk viscosity. As can be seen from the results of Appendix~\ref{app:vis}, at the background level, the pressure and bulk viscosity are degenerate and always appear in the form $p-3H\zeta$. It is this combination that we denoted by $p_{\text{eff}}$. However, when we incorporate fluctuations, the degeneracy will be broken. Note that the effective equation of motion for fluctuations, Eq.~(\ref{eq:eftsch}), has been obtained to $\order{\epsilon}$. At this order, the background fluid is effectively pressureless. As a result, in the equations for fluctuations of a viscous fluid we set $p_{\text{eff}}=0$ or $p=3H\zeta$ wherever these background quantities appear. 

We may define the {\it comoving overdensity} of the fluid, $\delta_{\rm eff}$, in terms of the right-hand side of the effective Poisson equation, Eq.~\eqref{eq:eftpoisson}:
\eq{
	\frac{\nabla^2\Phi_s}{a_s^2}=\frac{1}{2\Mpl^2}\rho_{\text{eff}} \, \delta_{\text{eff}}\,,	
}
where $\rho_{\text{eff}}$ is the effective energy density of the background in Eq.~\eqref{eq:rho_eff}, which, to ${\cal O}(\epsilon )$, is $\rho_{\text{eff}}=m|\bar{\psi}_s|^2$. This yields
\eq{
\delta_{\text{eff}}=\left(\frac{\dpsi_s}{\bar{\psi}_s} + \frac{\dpsi_s^*}{\bar{\psi}_s^*}\right) + i\frac{3H_s}{2m}\left(\frac{\dpsi_s}{\bar{\psi}_s} - \frac{\dpsi_s^*}{\bar{\psi}_s^*}\right)-\Phi_s\,.
\label{eq:deltaeffdef}
}
Note that although $\delta_{\rm eff}$ is constructed from complex quantities, the combination in Eq.~(\ref{eq:deltaeffdef}) remains real. Other fluid fluctuations, such as the effective fluctuations in density, velocity, and pressure, can also be derived by comparing the effective equations in Sec.~\ref{sec:EFT} with the fluid equations outlined in Appendix~\ref{app:vis}. This results in 
\eqa{
	\label{eq:drhoeff}
&\delta\rho_{\text{eff}}=m|\bar{\psi}_s|^2 \left( \frac{\delta\psi_s}{\bar{\psi}_s} + \frac{\delta\psi_s^*}{\bar{\psi} _s^*}-\Phi_s\right)\,,\\
\label{eq:dueff}
&\delta u_{\text{eff}}=\frac{1}{2mi}\left(\frac{\delta\psi_s}{\bar{\psi}_s} - \frac{\delta\psi_s^*}{\bar{\psi}_s^*}\right) + \frac{3H_s}{4m^2}\left(\frac{\delta\psi_s}{\bar{\psi}_s} + \frac{\delta\psi_s^*}{\bar{\psi}_s^*}\right) + \frac{1}{2mi}\frac{\nabla^2}{4m^2a_s^2}\left(\frac{\delta\psi_s}{\bar{\psi}_s} - \frac{\delta\psi_s^*}{\bar{\psi}_s^*}\right)\,,\\
&\delta
\label{eq:dpeff} p_{\text{eff}}=m|\bar{\psi} _s|^2 \left[ \left( -\frac{\nabla^2}{4m^2a_s^2} - \frac{\nabla^4}{8m^4a_s^4} + \frac{21H_s^2}{16m^2}\right)\left(\frac{\delta\psi_s}{\bar{\psi}_s}+\frac{\delta\psi_s^*}{\bar{\psi}_s^*} \right) - \frac{3iH_s\nabla^2}{8m^3a_s^2}\left(\frac{\delta\psi_s}{\bar{\psi}_s} - \frac{\delta\psi_s^*}{\bar{\psi}_s^*}\right)\right]
.}
Note that the above gauge-dependent variables are written in the Newtonian gauge. (See Appendix \ref{app:gauge} for relevant gauge transformations and corresponding expressions in the {\it time-averaged comoving} gauge.)

Using Eq.~\eqref{eq:deltaeffdef} and Eqs.~\eqref{eq:eftsch} and \eqref{eq:eftphidot} we can derive a second-order differential equation for $\delta_{\text{eff}}$, from which we can read new variables for the fluid, by making the analogy between the resulting equation and the standard second-order equation for an imperfect fluid outlined in  Appendix~\ref{app:vis} (see Eq.~\eqref{eq:ddot_fluidp0}). In this way, we obtain
\eq{
	\label{eq:eftdelta}
	\ddot{\delta}_{\text{eff}}+2H_s\dot{\delta}_{\text{eff}}-c^2_{\text{eff}}\frac{\nabla^2\delta_{\text{eff}}}{a^2}-\frac{\zeta_{\text{eff}}}{\rho_{\text{eff}}}\frac{\nabla^2\dot{\delta}_{\text{eff}}}{a_s^2}=\frac{\rho_{\text{eff}}}{2\Mpl^2}\delta_{\text{eff}}\,,
}
where we have defined the effective speed of sound by
\begin{eBox}
\eq{
	\label{eq:eftcs}
	c^2_{\text{eff}}=\frac{k^2}{4m^2a_s^2}-\frac{k^4}{8m^4a_s^4}+\frac{15}{16}\frac{H_s^2}{m^2}\,,
}
\end{eBox}
and the effective coefficient of bulk viscosity by
\begin{eBox}
\eq{
	\label{eq:eftzeta}
	\zeta_{\text{eff}}=-\frac{H_s}{2m^2}\rho_{\text{eff}}\,.
}
\end{eBox}
The bulk viscosity coefficient may be re-expressed in terms of another quantity with dimension of velocity \cite{Carrasco:2012cv}:
\ba 
c_{bv}^2 \equiv  \dfrac{H_s}{\rho_{\text{eff}}} \zeta_{\text{eff}}=-\dfrac{H_s^2}{2m^2},
\ea 
which may be compared with $c^2_{\text{eff}}$ in Eq.~\eqref{eq:eftcs}.
	
Some remarks are in order regarding these results. The leading term in the effective speed of sound in Eq.~\eqref{eq:eftcs} is well known \cite{Hwang:2009js,Poulin:2018dzj}. However, we find two additional contributions. The second term is a higher-order, momentum-dependent contribution. A similar term also appears in the analyses of Refs.~\cite{Hwang:2009js,Poulin:2018dzj} but with a different coefficient. (Our term is larger than the result of Refs.~\cite{Hwang:2009js,Poulin:2018dzj} by a factor of 2.) We trace this discrepancy to the fact that in Refs.~\cite{Hwang:2009js,Poulin:2018dzj}, the backreaction of the nonzero modes has been neglected, though they contribute at the same order. Note also that, despite the claims of Ref.~\cite{Hwang:2009js} (and also recently in Ref.~\cite{Lague:2020htq}), we do not expect these results to hold for arbitrary momentum, since the whole formalism breaks down as soon as the momentum of the field becomes comparable to its mass. However, in Minkowski spacetime, one can make a nonlocal field redefinition which yields an EFT that is nonperturbative in $\epsilon_k$ and holds for arbitrary momentum. Upon doing so, we confirm our coefficient, which is different from that obtained in Refs.~\cite{Hwang:2009js,Poulin:2018dzj}. (See Appendix~\ref{app:nonlocal} for details, where we also show that, thanks to the nonlocal field redefinition, the resulting sound speed approaches unity in the large-momentum limit, consistent with the expectation for a scalar field with canonical kinetic energy.) 

Finally, the last term in Eq.~\eqref{eq:eftcs}, which is a nontrivial, momentum-independent contribution from the background evolution, is new and has not been identified in previous analyses. This term shows that all field fluctuations experience an effective sound speed, even modes that are well beyond the horizon. Note that since the effective sound speed is always positive, it suppresses structure formation. Explicit investigation of this effect would be interesting but is beyond the scope of this paper.

Besides the sound speed, we have also derived the coefficient of the bulk viscosity in Eq.~\eqref{eq:eftzeta}, which has also been omitted in previous analyses. In principle one could have anticipated that such a term would appear from an EFT perspective, since it is consistent with the symmetries of the system. Note that this coefficient is \emph{negative}. From the second law of thermodynamics it can be shown that the coefficient of bulk viscosity of an isolated, imperfect fluid in thermal equilibrium must be positive \cite{Landau:1959}. However, our nonrelativistic system is {\it not} an isolated system: it exchanges energy with the relativistic sector of the field/fluid, and our effective description remains ignorant about the latter. Therefore, the sign in Eq.~(\ref{eq:eftzeta}) is not inconsistent with the second law of thermodynamics.\footnote{Taking both sectors into account simultaneously must result in a non-negative viscosity coefficient, and, indeed, a canonically normalized (relativistic) scalar field, re-expressed in terms of fluid dynamics, shows no viscocity. In a similar way, the usual energy conditions, such as the strong, weak, and dominant energy conditions, need not hold in a low-energy EFT, even if the underlying (relativistic) theory obeys them \cite{Creminelli:2006xe,Azhar:2018nol}.} A negative bulk viscosity would lead to an enhancement in the development of structures, competing with the positive sound speed. It would be interesting to investigate which contribution dominates, and whether the balance depends on time or length-scale. 

\section{Comparison with numerical solutions}\label{sec:num}
In this section we show that the effective description for the slow modes, as well as the nontrivial expressions for the nonzero modes obtained in Sec.~\ref{sec:EFT}, are consistent with the smearing of the exact solutions by the method discussed in Sec.~\ref{sec:mexp}. We also show that the EFT is able to construct the full oscillating solution with good accuracy, while our effective equations are well-behaved with no oscillatory terms.

In Sec.~\ref{sec:mexp} we adopted the window function $W(t)$ of Eq.~\eqref{eq:w} for obtaining different modes of each variable. It was an appropriate choice for a mathematically rigorous formulation of the smearing procedure and for the exact mode decomposition. However, it is impractical for numerical purposes. The amplitude of the sinc function does not fall off sufficiently rapidly and, further, the naive truncation of the function $W (t)$ beyond the range of the simulation time results in the Gibbs phenomenon at discontinuities after a (numerical) Fourier transformation \cite{Proakis}. To efficiently circumvent these issues in our numerical smearing (as is well-known in the field of signal processing) we replace $W(t)$ with $W(t) {\cal B}(t/T)$, where ${\cal B}(t)$ is the {\it Blackman} window function \cite{Proakis} and $T$ is the total time of the simulation.
Fig.~\ref{fig:modes} depicts the results of this smearing procedure for the slow mode as well as the leading nonzero modes for the exact numeric solutions of $\psi$ and $\delta \psi$ (i.e. the solutions of Eqs.~\eqref{eq:bckg1}, \eqref{eq:bckg2}, \eqref{eq:linpert1} and \eqref{eq:linpert2}). In the same figure, we compare these results with the predictions of our EFT, the solutions of Eqs.~\eqref{eq:EFTbck1}, \eqref{eq:EFTbck2}, \eqref{eq:eftsch}, \eqref{eq:eftphidot}, and \eqref{eq:eftpoisson}, along with the nonzero modes obtained in Eqs.~\eqref{eq:psinu1}, \eqref{eq:psinu2}, and \eqref{eq:Phinu2} in terms of the slow modes. This confirms that our EFT predictions match the numerical smearing with good accuracy. Fig.~\ref{fig:modes} also shows the hierarchy between nonzero modes among themselves and with the slow mode as predicted by the EFT.

As we saw in Sec.~\ref{sec:EFT}, all the variables can be represented by a mode expansion. For instance for the background variable $\bar{\psi} (t)$, to working order, we can write
\eq{
\label{eq:psiexp}
\bar{\psi} (t)= \bar{\psi} _s + \bar{\psi}_2 e^{2imt} + \bar{\psi}_4 e^{4imt}+ \bar{\psi}_{-2} e^{-2imt}+\dots\,,
}
while similar expressions hold for the other variables. 
\fg{
\centering
\includegraphics[width=0.45\textwidth]{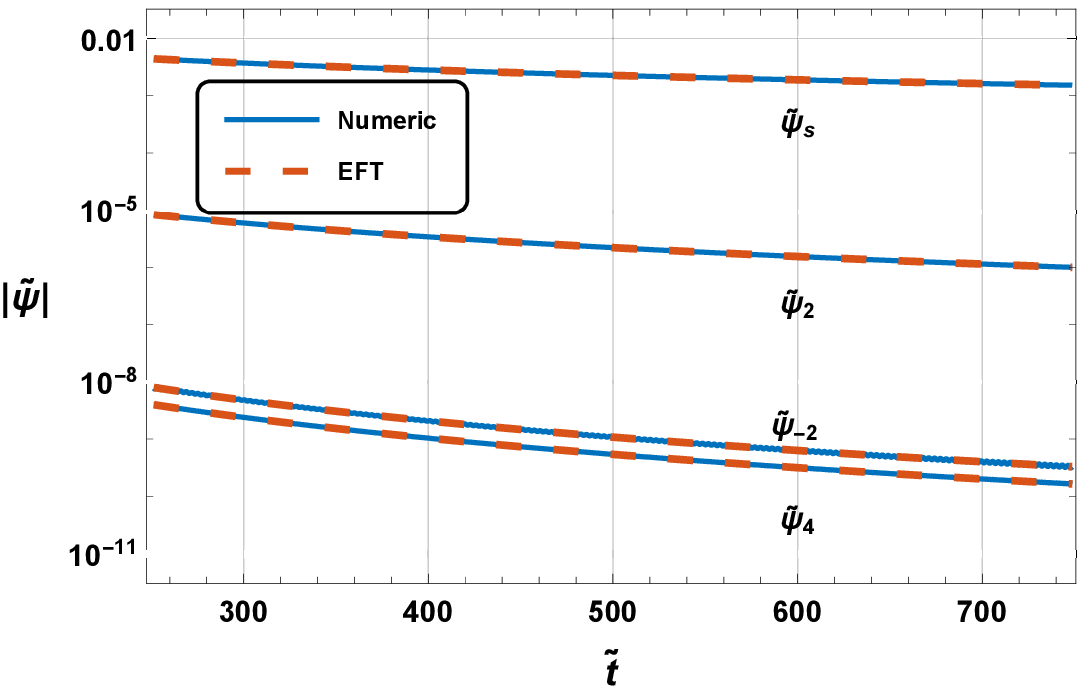}
\hskip 1cm
\includegraphics[width=0.45\textwidth]{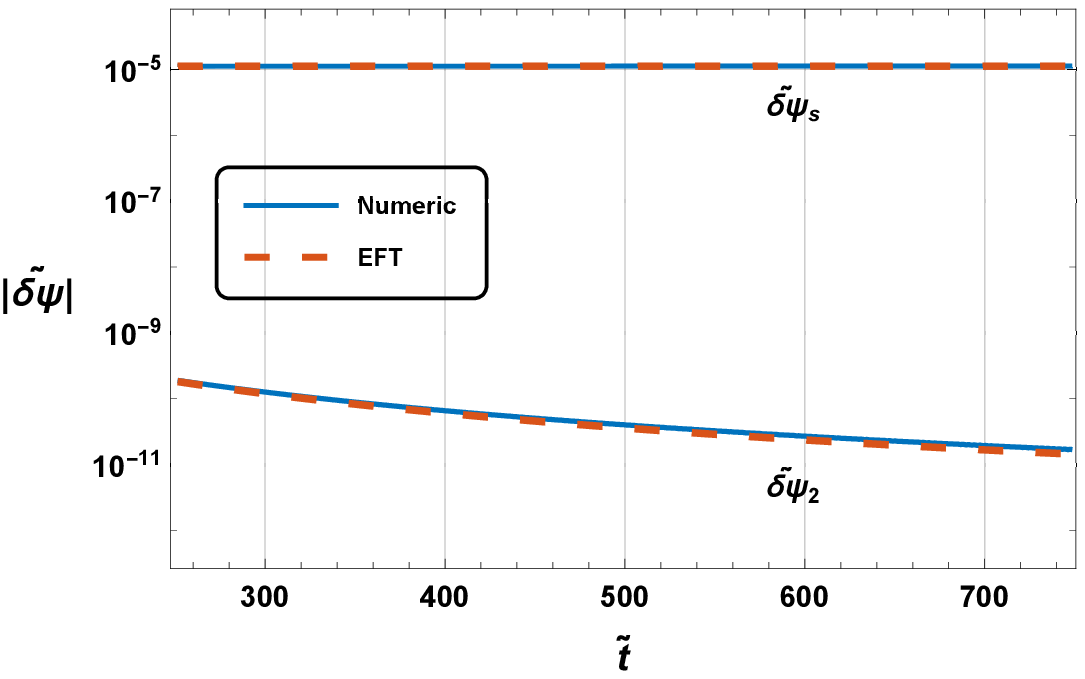}
\caption{Absolute values of the slow mode as well as the leading nonzero modes for $\cpsi(t)$ (left) and $\cdpsi(t)$ (right) obtained by smearing the exact numerical solutions (solid, blue curve) compared with the predictions of our EFT (dashed, orange curve).  We have set $H(0)/m=0.06$, $k/m=0.02$ and  $a_s(0)=1$ which, from the EFT, implies $a(0)=0.998$. The initial conditions for the slow mode are $\cpsi_s(0)=0.1+0.025i$ and $\cdpsi_s(0)=(1-1.75i)\times10^{-6}$  which, from the EFT, imply $\cpsi(0)=0.098+0.020i$ and  $\cdpsi(0)=(0.326-1.61i)\times10^{-6}$ for the exact equations. The horizon crossing for the (randomly chosen) mode occurs around $\tilde{t}=25$.}
\label{fig:modes}
}
As mentioned earlier, since the nonzero modes are also expressed in terms of the slow mode, solving the obtained effective equations for the slow mode allows us to construct the full solution by relations like Eq.~\eqref{eq:psiexp} order by order in our perturbation theory. In Figs.~\ref{fig:ReImpsi} and \ref{fig:ReImdpsi} we compare the exact solutions of $\bar{\psi}$ and $\delta \psi$ with the solution for the slow mode in our EFT (labeled ``EFT, slow mode''), the full solution constructed out of our EFT (labeled``EFT, constructed''), and the ``naive'' theory (where all oscillatory terms are simply neglected in the equations of motion, as is typically done in analyses of axion-like dark matter models). It is evident from these plots that the naive theory can cause notable error, while the EFT solution follows the exact solution quite reliably. We, however, warn that the amount of deviation is sensitive to the small parameters based on which our EFT is constructed. (Here, for example, we take $H/m=0.06$ and $k/am=0.02$ at the initial time of the simulation, a somewhat random choice.) An interesting investigation, which we will leave for future work, would be to study the error that one  encounters by using the naive theory in a realistic situation where, e.g., the other components of matter in the universe are taken into account and the parameter space is chosen based on observational constraints.

\fg{
\includegraphics[width=0.47\textwidth]{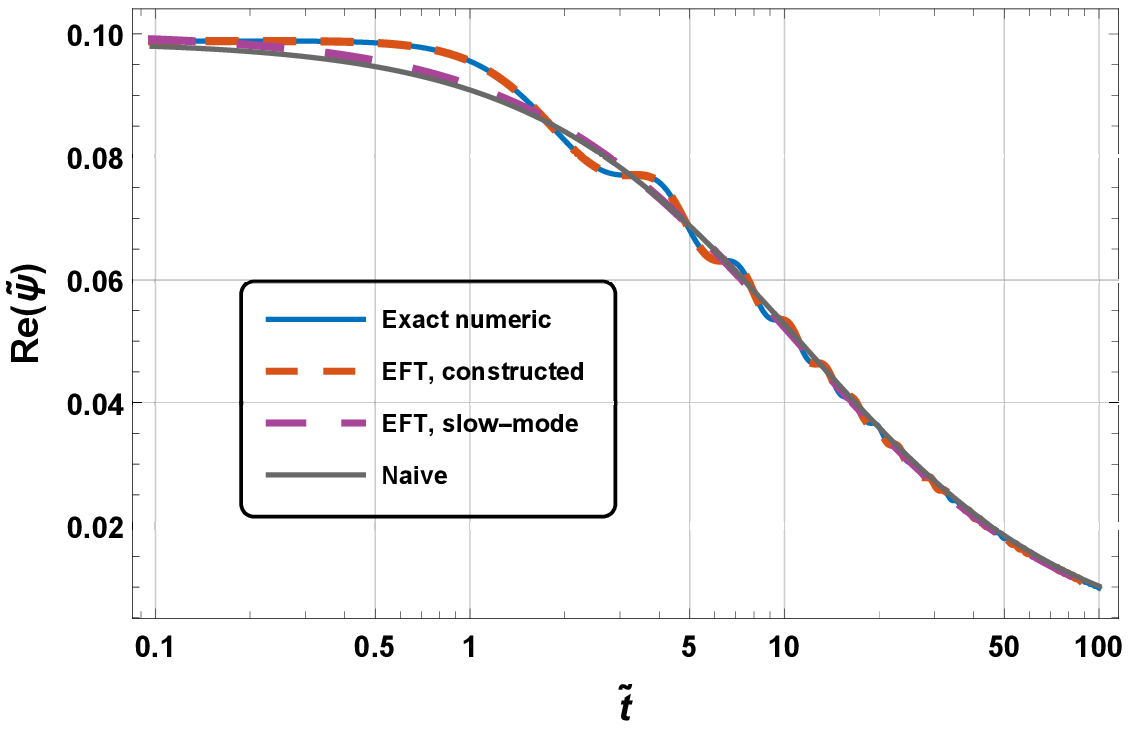}
\hspace{5mm}
\includegraphics[width=0.48\textwidth]{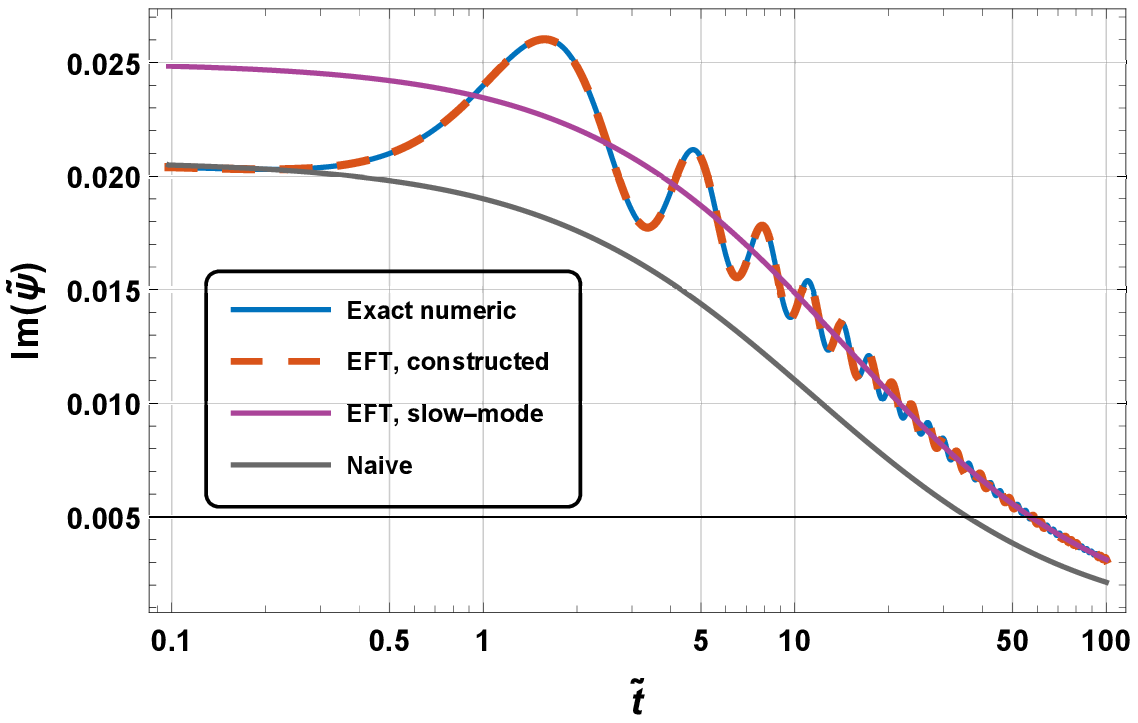}
\caption{Comparison of the real (left) and imaginary part (right) of $\cpsi$ between the exact numerical results, the full solution constructed out of EFT, the slow mode, and the naive solutions. The choice of parameters and initial conditions are the same as Fig.~\ref{fig:modes}.}
\label{fig:ReImpsi}
}
\fg{
\includegraphics[width=0.48\textwidth]{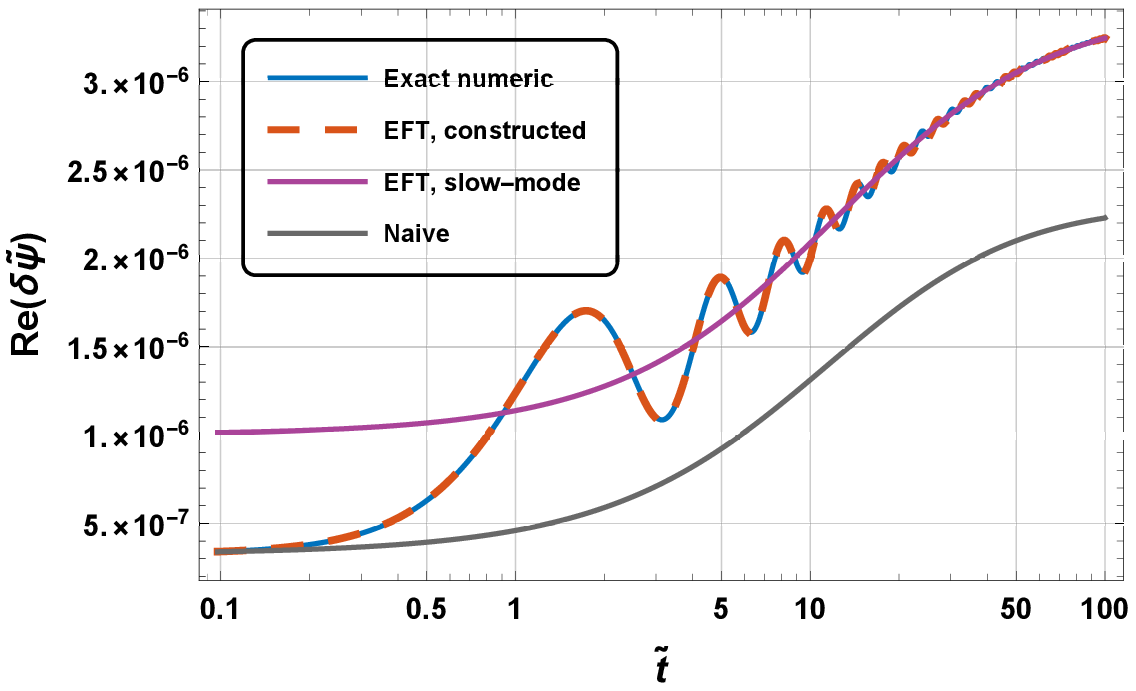}
\hspace{5mm}
\includegraphics[width=0.47\textwidth]{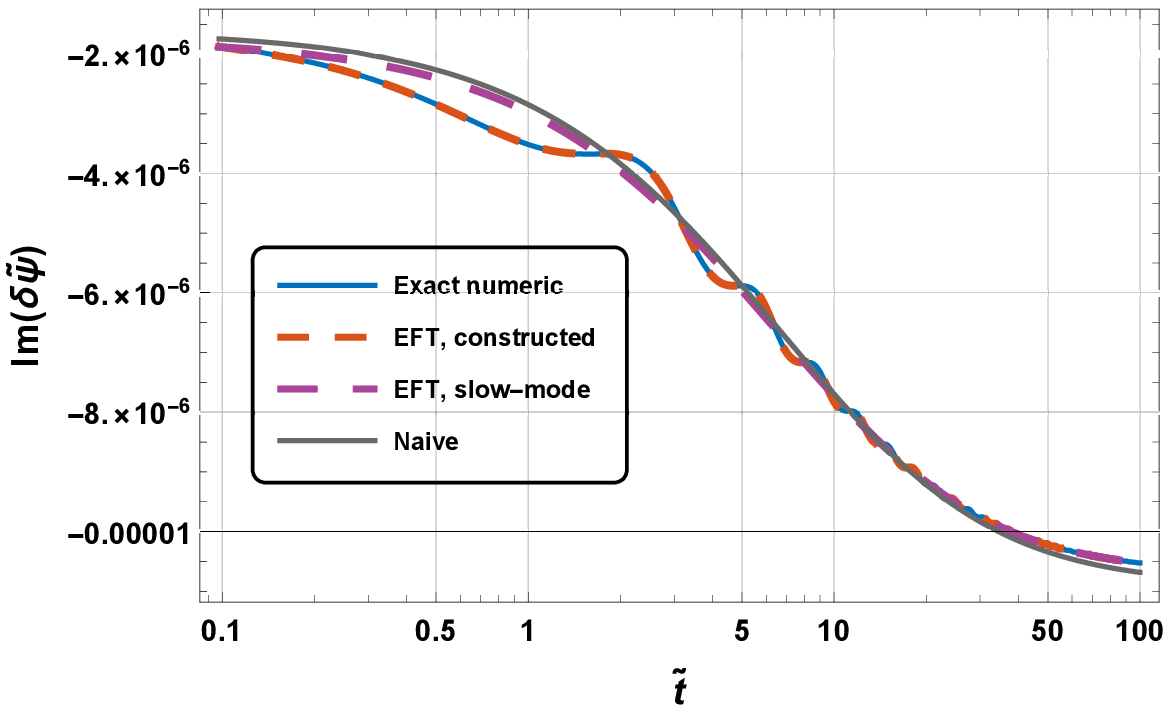}
\caption{Comparison of the real (left) and imaginary part (right) of $\cdpsi$ between the exact numerical results, the full solution constructed out of EFT, the slow mode, and the naive solutions. The choice of parameters and initial conditions are the same as Fig.~\ref{fig:modes}.}
\label{fig:ReImdpsi}
}

\fg{
\includegraphics[width=0.48\textwidth]{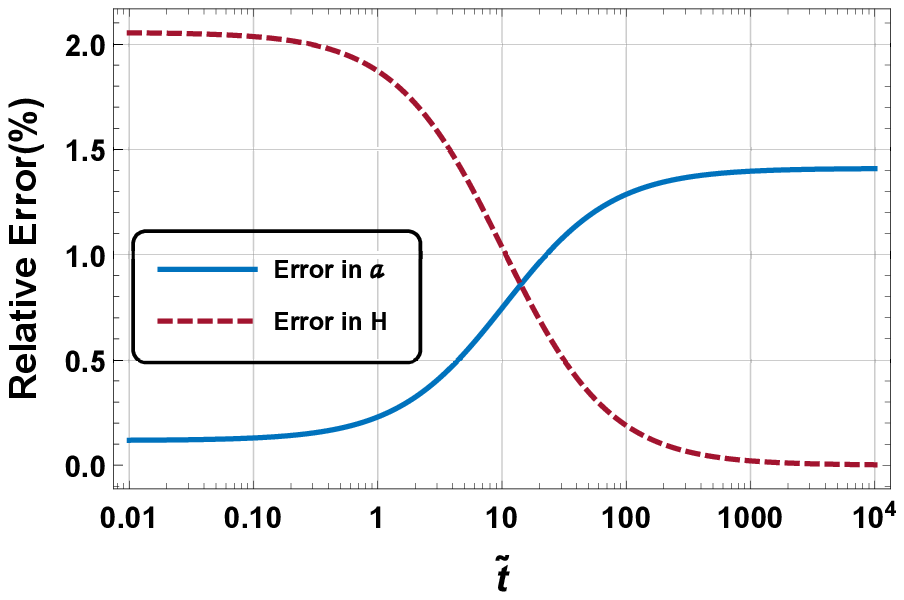}
\hspace{5mm}
\includegraphics[width=0.47\textwidth]{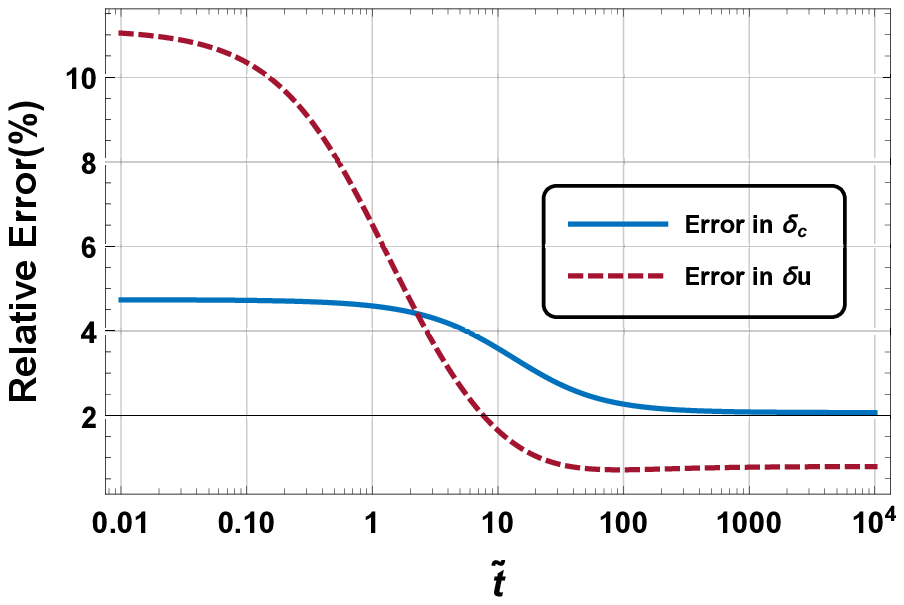}
\caption{Relative error in variables $H$, $a$, $\delta_c$, and $\delta u$ for the naive theory compared with the EFT. The choice of parameters and initial conditions are the same as Fig.~\ref{fig:modes}.}
\label{fig:H}
}

A remark regarding the choice of initial conditions for comparison is in order. For the naive theory (which simply neglects any rapidly oscillating contribution) we have no option other than choosing the initial conditions to be the same as the exact theory. In the EFT, on the other hand, we have perturbative access to the full solution, as just described above, so that we can match the initial conditions of the full solution constructed out of EFT to the ones chosen for the exact solution.{\footnote{In fact, we found it more convenient to fix the initial conditions for the EFT first and then obtain the corresponding initial conditions for the exact solution. Note, however, that the inverse procedure (which is more realistic) is also possible.}} That is why in Figs.~\ref{fig:ReImpsi} and \ref{fig:ReImdpsi} the initial condition for the slow mode is different from the exact solution (since nonzero modes contribute to the initial conditions as well). Note that this is the natural choice of the initial conditions for our EFT, suggested by the EFT itself.  

Finally, having confirmed the validity of our EFT, we can consider the results of the EFT to be a sufficiently accurate description of the dynamics of the system, and then compare it with the naive theory in variables that are of observational interest. In Fig.~\ref{fig:H} we make this comparison for the Hubble parameter, the scale factor, the density contrast, and the velocity potential. Depending on the parameters and initial conditions, the error seems to be in fact observable.  The quantification of that statement in realistic situations, however, is beyond the scope of this paper and will be studied elsewhere.

\section{Summary and Outlook}\label{sec:con}
In this paper we have obtained an EFT for a massive, nonrelativistic scalar field in an expanding background by systematically integrating out rapidly oscillating modes. We applied our formalism to spatially homogeneous quantities as well as to spatially varying fluctuations (working to linear order in the fluctuations), but the same methods can be employed for studying any system in which there are fast oscillations while the physically interesting variables are slowly varying in time. For the sake of convenient access, we summarize our main results here. The effective equations governing the dynamics of the system (in an expanding massive-field-dominated universe) are 
\eqa{
i\dot{\bar{\psi}}_s +\frac{3i}{2}H_s \bar{\psi}_s + \frac{9}{16}\left[1+\frac{iH_s}{2m}\right] \frac{|\bar{\psi}_s|^2}{\Mpl^2} \bar{\psi}_s \simeq 0\, , \qquad 
3\Mpl^2H_s^2\simeq m|\bar{\psi}_s|^2 + \frac{3}{32\Mpl^2}|\bar{\psi}_s|^4\, ,
} 
for the background variables (up to ${\cal O}(\epsilon^2)$), and

\eqa{
&i\dot{\dpsi_s}+\frac{3i}{2}H_s\dpsi_s + \left( 1+\frac{\nabla^2}{4m^2a_s^2}\right) \frac{\nabla^2\dpsi_s}{2ma_s^2} - \left(1-\frac{2iH_s}{m}\right) m\bar{\psi}_s \Phi_s \nonumber \\
&\quad\quad\quad\quad + \frac{9|\bar{\psi}_s|^2}{8\Mpl^2}\dpsi_s - \frac{7\bar{\psi}_s^2}{16\Mpl^2}\dpsi_s^* \simeq 0\, ,\\
&\dot{\Phi}_s + H_s\Phi_s + \frac{1}{4\Mpl^2} \left[i\left( 1+\frac{\nabla^2}{4m^2a_s^2}\right) (\bar{\psi}_s \dpsi_s^* - \bar{\psi}_s^* \dpsi_s) +\frac{3H_s}{2m}(\bar{\psi}_s \dpsi_s^* + \bar{\psi}_s^* \dpsi_s) \right]\simeq 0\,,\\ \nonumber \\
&\left[\frac{\nabla^2}{a_s^2}+\frac{3}{2}H_s^2\right]\Phi_s\simeq  \frac{m}{2\Mpl^2} (\bar{\psi}^*_s\dpsi_s + \bar{\psi}_s \dpsi^*_s) + \frac{3iH_s}{4\Mpl^2} (\bar{\psi}^*_s \dpsi_s - \bar{\psi}_s\dpsi^*_s)\,,
} 
\\
for fluctuations (up to ${\cal O}(\epsilon)$), where, as a reminder, we parameterize the field as $\psi (t, {\bf x}) = \bar{\psi} (t) + \delta \psi (t, {\bf x})$. To leading order, these equations correspond to the already well-known Schr\"odinger and Schr\"odinger-Poisson equations for the background evolution and for the fluctuations, respectively. However, in our EFT, we have obtained nontrivial corrections as a result of integrating out the nonzero, rapidly oscillating modes (rather than neglecting them).  Furthermore, we have also interpreted the results more intuitively by describing the system as an effective (imperfect) fluid. To fully describe the system, to working order, we identify a nonzero effective pressure as well as  an effective sound speed and a bulk viscosity as follows:
\eq{
p_{\text{eff}}=\frac{9}{32 \Mpl^2}|\bar{\psi}_s|^4\,,\qquad c^2_{\text{eff}}=\frac{k^2}{4m^2a_s^2}-\frac{k^4}{8m^4a_s^4}+\frac{15}{16}\frac{H_s^2}{m^2}\,,\qquad \zeta_{\text{eff}}=-\frac{H_s}{2m^2}\rho_{\text{eff}}\,.
}
Note that the pressure and the bulk viscosity were missing in all previous analyses. Furthermore, the second term in the sound speed has a different numerical prefactor, compared to other results in the literature (see e.g., Refs.~\cite{Hwang:2009js,Poulin:2018dzj}). The discrepancy seems to be due to an extra contribution as a result of nontrivial effects of oscillatory (nonzero) modes, neglected in other studies. The last term in the sound speed was also missing in previous analyses. The size of the error arising by neglecting these terms requires further investigation in realistic situations, which is beyond the scope of this paper.

The derived EFT is interesting from both a theoretical and a practical point of view. On the theoretical side we can see that gravity induces an effective, attractive self-interaction in a free scalar field theory, which manifests as nontrivial pressure, sound speed, and viscosity. These effects can be important for the background evolution as well as for the growth of overdensities. Note that the effective sound speed is positive while the viscosity is negative, so that they act in opposite directions: the former suppresses the growth of overdensities while the latter tends to enhance it. It would be interesting to investigate in which situations the various variables win, and how their incorporation changes the results compared to the naive theory. We leave such a study for future work. 

From a practical point of view, as we saw in Sec.~\ref{sec:num}, whereas the naive treatment can deviate substantially from the exact results and hence may cause error in interpreting observations, our EFT remains quantitatively reliable. In addition, our method paves the way for efficiently obtaining accurate solutions (including oscillatory behavior) without having to solve the exact equations (which are expected to be stiff due to rapid oscillations). Note that simulating the exact theory numerically requires time increments $\Delta t<1/m$ in order to accurately capture effects of the oscillating terms, whereas the corresponding equations within our EFT remain well-behaved, with no rapidly oscillating terms, so that it is sufficient to use $\Delta t<{\text{min}}(1/H,a/k)$; roughly speaking, this yields an $\order{1/\epsilon}$ increase in efficiency. The full solution can then be constructed order by order using the mode decomposition outlined in Sec.~\ref{sec:EFT} with no difficulty. Such a method for solving differential equations containing rapidly oscillating terms can have interesting applications in much broader situations of scientific interest. 

There are a number of different directions --- besides the ones already mentioned --- that we would like to explore in future work. Such studies include the application of our EFT to predictions of possible impacts of ultra-light dark matter models on the CMB and other observations; the EFT in the nonlinear regime and the corresponding corrections to structure formation (cf. Ref.~\cite{Musoke:2019ima}) as well as the dynamics of celestial objects as they move through dark matter halos; and the extension of our EFT to the situation in which the nonrelativistic field is coupled to another (perhaps relativistic) dynamical field. 


\acknowledgments

We are grateful to Xingang Chen, Mohammad Ali Gorji, Mahdiyar Noorbala, Katelin Schutz, and Tracy Slatyer for helpful discussions. B.S. thanks YITP at Kyoto University for hospitality during the time this work was in progress. Portions of this work were conducted in MIT's Center for Theoretical Physics and supported in part by the U.S. Department of Energy under Contract No.~DE-SC0012567.

\appendix

\section{The Hamiltonian}\label{app:Hamiltonian}
In this appendix we will find the Hamiltonian for the evolution of $\psi$ in a general curved background. Since the gravity sector is standard and we have not performed any redefinition for the metric we will assume the gravity sector as a fixed background with given time evolution. Standard treatment of the Hamiltonian for the metric can be found for example in Ref.~\cite{Poisson:2009pwt}. Here, instead of deriving the Hamiltonian from the rather involved Lagrangian for the $\psi$ field in Eq.~\eqref{eq:Lpsipsistar}, we first work out the Hamiltonian of the Lagrangian in Eq.~\eqref{eq:phichilag} in terms of $\phi$ and the auxiliary field $\chi$ and then perform the field redefinition, via a canonical transformation, to derive a Hamiltonian for $\psi$ and $\psi^*$. Recall that after removing $\dot \phi$ in favor of a new variable $\chi$ and integrating out a nondynamical field, we obtained the Lagrangian in Eq.~\eqref{eq:phichilag}, which we repeat here:
\eq{
	\label{eq:phichilag:repeat}
\mathcal{L}(\phi,\chi)=-\frac{1}{2}\sqrt{-g}\left[-g^{00}\chi^2+2(g^{00}\chi+g^{0i}\p_i\phi)\dot{\phi}+g^{ij}\p_i\phi\p_j\phi+2V\right]\,. 
}
Here, we have again written the potential term in its general form $V(\phi)$ for the scalar field which can contain both the mass term and the interaction. Before computing the Hamiltonian it is instructive to count the number of degrees of freedom in our theory. Since the metric is assumed to be fixed, in the original scalar field theory we had only two degrees of freedom, $(\phi,\pi_\phi)$. In Eq.~(\ref{eq:phichilag:repeat}), we introduced a different Lagrangian which is completely equivalent, in the sense that it leads to the same equations of motion. Naively, the Lagrangian in Eq.~(\ref{eq:phichilag:repeat}) appears to involve two fields, $\phi$ and $\chi$, so in principle it may have four degrees of freedom, $(\phi,\chi,\pi_\phi,\pi_\chi)$. However, the new theory is a constrained system \cite{Henneaux:1992ig,Weinberg:1995mt}. To see this we compute the conjugate momenta 
\eqa{
	\pi_\phi&=\fdv{\mathcal{L}}{\dot{\phi}}=-\sqrt{-g}(g^{00}\chi+g^{0i}\p_i\phi)\\
	\pi_\chi&=\fdv{\mathcal{L}}{\dot{\chi}}=0\,.
}
We therefore see that we cannot solve for $\dot{\phi}$ and $\dot{\chi}$ in terms of phase space variables, that is, the Lagrangian is degenerate. The dynamics in phase space are constrained to a part of the phase space specified by primary constraints $C_i=0$ for $i\in\{1,2\}$, where
\eqa{
\label{eq:c1}
C_1&=\pi_\chi\\
\label{eq:c2}
C_2&=\pi_\phi+\sqrt{-g}(g^{00}\chi+g^{0i}\p_i\phi)\, .
}  
These two constraints reduce the number of degrees of freedom to $4-2=2$, which is consistent with the original theory. The Hamiltonian density can then be computed from the standard procedure for a constrained system:
\eq{
	\label{eq:ttham}
	\mathcal{H}_T=u_1C_1+u_2C_2+\mathcal{H}\,,
}
with
\eq{
	\label{eq:ham1}
	\mathcal{H}=\frac{1}{2}\sqrt{-g}\left[-g^{00}\chi^2+g^{ij}\p_i\phi\p_j\phi+2V\right]\,,	
}
and $u_1$ and $u_2$ are unspecified functions. Note that the dynamics is controlled by the total Hamiltonian $\mathcal{H}_T$, and we are not allowed to impose constraints on the total Hamiltonian before computating the relevant Poisson brackets. The constraints must be preserved in time; we therefore have the following set of equations describing the system:
\ba 
\label{eq:poisson_brackets}
\nonumber
\dot Q_i(\bfx) =\{Q_i(\bfx) , H_T \}\, , \> \dot P_i(\bfx)  = \{P_i(\bfx) , H_T\}, \> \dot C_i =\pdv{C_i}{t}+\{C_i ,H_T\}=0  , \> \text{with} \, \, i=1,2
\\
\ea 
where $Q_1=\phi, \, Q_2=\chi $, $P_i$ are their corresponding conjugate momenta, and $H_T=\int\dd[3]{x}\mathcal{H}_T$ is the total Hamiltonian. The first term on the right-hand side of the equation for $\dot C_i$ is due to the fact that the metric components (and hence constraints) can explicitly depend on time. Note that the Poisson brackets must be understood as operators in terms of functional derivatives, i.e. 
\ba 
\{f(\bfx),g(\bfx')\} \equiv \sum_{i=1,2} \, \int d^3\bfx'' \left[ \dfrac{\delta f(\bfx)}{\delta Q_i(\bfx'')} \dfrac{\delta g(\bfx')}{\delta P_i(\bfx'')}-
 \dfrac{\delta f(\bfx)}{\delta P_i(\bfx'')} \dfrac{\delta g(\bfx')}{\delta Q_i(\bfx'')} \right].
\ea 
From the set of equations in Eq.~\eqref{eq:poisson_brackets}, we find $u_1=\dot \chi$ and $u_2=\dot \phi$ as well as the Klein-Gordon equation for the original field $\phi$. 

An alternative approach which can lead to a simplified Hamiltonian formulation in a constrained system is to use the Dirac formalism \cite{Henneaux:1992ig}. To employ this method, we first note that the Poisson bracket of constraints is nonzero. Let us then define
\eq{
	\Delta_{ij}(\bfx, \bfx')\equiv \{C_i(\bfx),C_j(\bfx')\}\, .
}  
It can be shown that
\eq{
\Delta=
\begin{bmatrix}
0&\quad -\mathcal{N}\\
\mathcal{N}& \, \quad \left( \mathcal{N}^i\p_i-\mathcal{N}^i{}'\p'_i \right)
\end{bmatrix} \times \delta^3(\bfx-\bfx')\,,
}
where we have defined $\mathcal{N}\equiv \sqrt{-g}g^{00}$ and $\mathcal{N}^i \equiv \sqrt{-g}g^{0i}$. A prime over $\mathcal{N}^i{}'$ denotes that the argument is evaluated at $\bfx'$, while $\p'_i$ denotes differentiation with respect to $x_i'$.
Note that the bracket of $C_2$ with itself is nonzero due to the presence of the spatial derivative terms. With the help of the inverse matrix $\Delta^{ij}$, we can construct Dirac brackets defined by
\eq{
	\{f(\bfx),g(\bfx')\}_D=\{f(\bfx),g(\bfx')\}-\int d^3 y\, d^3y'\, \{f(\bfx),C_i({\bf y})\}\Delta^{ij}({\bf y,y'})\{C_j({\bf y'}),g(\bfx')\}\,.
}
The inverse can be computed to be

\eq{
	\Delta^{-1}=\dfrac{1}{\mathcal{N}\,\mathcal{N}'}
	\begin{bmatrix}
		\left( \mathcal{N}^i\p_i-\mathcal{N}^i{}'\p'_i \right) &\qquad \mathcal{N}' \\
		-\mathcal{N}' &\qquad 0
	\end{bmatrix}
\times \delta^3(\bfx-\bfx')
\,,
}
where by inverse we mean
\eq{
	\sum_{k}\int\dd[3]{ y}\Delta_{ik}({\bf x,y})\Delta^{-1}_{kj}({\bf y,x'})=\delta_{ij}\delta^3({\bf{x-x'}})\,.
}
Using the Dirac brackets allows us to use the simplified Hamiltonian, in which the constraint terms are removed. In other words, we simply use the Hamiltonian $\mathcal{H}$ in Eq.~\eqref{eq:ham1}, with the price paid that the Poisson brackets are replaced with Dirac ones. Two relevant brackets are
\eq{
	\label{eq:phichi}
	\{\phi,\chi\}_D=-\frac{1}{\mathcal{N}}\delta^3({\bf{x-x'}})\,,
}
and
\eq{
	\label{eq:chichi}
	\{\chi,\chi\}_D=\frac{1}{\mathcal{N}\,\mathcal{N}'}\left( \mathcal{N}^i\p_i-\mathcal{N}^i{}'\p'_i \right)\delta^3({\bf{x-x'}}).
}
One can then see that $\dot Q_i =\{Q_i, H\}_D$ with $H=\int \calH d^3x $ gives the correct equations of motion.

The next step is to express everything in terms of $\psi$ and $\psi^*$ with the help of the field redefinition of Eq.~\eqref{eq:cantr1}. We must add to Eq.~\eqref{eq:cantr1} a suitable transformation for their conjugate momenta so the transformation is guaranteed to be a canonical transformation. We use the standard procedure of constructing a generating function \cite{Goldstein}. Let us consider a generating function of old momenta $\pi_\phi$ and $\pi_\chi$ and new variables $\psi$ and $\psi^*$ of the form $F_3(\pi_\phi,\pi_\chi,\psi,\psi^*)$. Then we must have
\eq{
\phi=-\fdv{F_3}{\pi_\phi}\,,\qquad\chi=-\fdv{F_3}{\pi_\chi}\,,
}
which, by using Eq.~\eqref{eq:cantr1}, we can find $F_3$ to be
\eq{
F_3=\int\dd[3]{x}\left[-\frac{\pi_\phi}{\sqrt{2m}}(e^{-imt}\psi+e^{imt}\psi^*)+i\sqrt{\frac{m}{2}}\pi_\chi(e^{-imt}\psi-e^{imt}\psi^*)\right]\,.
}
The Hamiltonian for $\psi$ and $\psi^*$ is then 
\eq{
\label{eq:H}
\mathcal{H}_\psi=\mathcal{H}(\psi,\psi^*)+\pdv{\mathcal{F}_3}{t}\,,
}
where the first term is the Hamiltonian of Eq.~\eqref{eq:ham1} expressed in terms of $\psi$ and $\psi^*$, which takes form
\eq{
\spl{
\mathcal{H}=\frac{\sqrt{-g}}{2}\bigg[&-mg^{00}\psi^*\psi+\frac{g^{ij}}{2m}(\p_i\psi\p_j\psi^*+\p_i\psi^*\p_j\psi)+2V
\\
&+\left\{\frac{e^{-2imt}}{2m}\left(g^{00}m^2\psi^2+g^{ij}\p_i\psi\p_j\psi\right)+\text{c.c.}\right\}\bigg]\,,
}
}
and the second term is a partial time derivative of the density $\mathcal{F}_3$, defined by $F_3=\int\dd[3]{x}\mathcal{F}_3$, which takes the form
\eq{
\pdv{\mathcal{F}_3}{t}=-\frac{\sqrt{-g}}{2}\bigg[mg^{00}(e^{-imt}\psi-e^{imt}\psi^*)^2+ig^{0i}(e^{-imt}\psi-e^{imt}\psi^*)\p_i(e^{-imt}\psi+e^{imt}\psi^*)\bigg]\,,
}
where we have imposed the constraints of Eqs.~\eqref{eq:c1} and \eqref{eq:c2} to eliminate old momenta at the level of the Hamiltonian. Therefore, we have
\eq{
\spl{
\label{eq:hpsi}
\mathcal{H}_\psi=\frac{\sqrt{-g}}{2}\bigg[&mg^{00}\psi^*\psi+\frac{g^{ij}}{2m}(\p_i\psi\p_j\psi^*+\p_i\psi^*\p_j\psi)-ig^{0i}(\psi\p_i\psi^*-\psi^*\p_i\psi)+2V\\
&+\left(\frac{e^{-2imt}}{2m}\left(-m^2g^{00}\psi^2-2img^{0i}\psi\p_i\psi+g^{ij}\p_i\psi\p_j\psi\right)+\text{c.c.}\right)\bigg]\,.
}
}
The final step is to find consistent Dirac brackets for $\psi$ and $\psi^*$. From the inverse transformation we have
\eq{
	\psi=\sqrt{\frac{m}{2}}e^{imt}\left(\phi+\frac{i}{m}\chi\right)\,,
} 
and using Eqs.~\eqref{eq:phichi} and \eqref{eq:chichi} it is easy to show that
\eq{
\{\psi,\psi^*\}_D= \left(\frac{i}{\mathcal{N}}+\frac{1}{2m \mathcal{N}\,\mathcal{N}'} \left(\mathcal{N}^i\p_i-\mathcal{N}^i{}'\p'_i\right)\right)\delta^3({\bf{x-x'}})\,,
} 
and
\eq{
\{\psi,\psi\}_D=-\frac{e^{2imt}}{2m \mathcal{N}\,\mathcal{N}'} \left(\mathcal{N}^i\p_i-\mathcal{N}^i{}'\p'_i \right)\delta^3({\bf{x-x'}})\,.
}
It is then simple to check that the relation $\dot \psi = \{\psi, H_\psi\}_D$ (and its complex conjugate) leads to the correct dynamics for the system, where $H_\psi=\int\dd[3]{x}\mathcal{H}_\psi$ is the Hamiltonian with the Hamiltonian density given in Eq.~\eqref{eq:hpsi}.

\section{Higher-order terms for fluctuations}\label{app:linhigher}
In Sec.~\ref{sec:eftpert}, we have only shown nonzero modes up to order $n=2$ for the fluctuations $\delta \psi (t, {\bf x})$ and $\Phi (t, {\bf x})$. For completeness, we present the general form to arbitrary order and also derive equations up to order $n=3$. From Eq.~\eqref{eq:Phinu} we can deduce for nonzero modes
\begin{equation}
\begin{split}
\Phi_\nu^{(n)}=&-\frac{1}{i\nu}\bigg[\Phi_\nu^{(n-1)'}+\cH_s\Phi_\nu^{(n-1)}+\cH_\nu^{(n-1)}\Phi_s+\sum_{\ell=1}^{n-1}\sum_{\alpha\neq0,\nu}\cH_\alpha^{(\ell)}\Phi_{\nu-\alpha}^{(n-\ell)}\bigg]\\
&-\frac{1}{4\nu}\bigg[\cpsi_s\cdpsi_s\delta_{\nu,-2}\delta_{n,2}-\cpsi_s^*\cdpsi_s^*\delta_{\nu,2}\delta_{n,2}+(1-\delta_{\nu,-2})(\cpsi_s\cdpsi_{\nu+2}^{(n-1)}+\cpsi_{\nu+2}^{(n-1)}\cdpsi_s)\\
&\hspace{12mm}-(1-\delta_{\nu,2})(\cpsi_s^*\cdpsi_{2-\nu}^{(n-1)}{}^*+\cpsi_{2-\nu}^{(n-1)}{}^*\cdpsi_s^*)+\cpsi_s\cdpsi_{-\nu}^{(n-1)}{}^*+\cpsi_\nu^{(n-1)}\cdpsi_s^*\\
&\hspace{12mm}-\cpsi_s^*\cdpsi_{\nu}^{(n-1)}-\cpsi_{-\nu}^{(n-1)}{}^*\cdpsi_s+\sum_{\ell=1}^{n-1}\sum_{\alpha\neq0,\nu}(\cpsi_{\alpha}^{(\ell)}\cdpsi_{\alpha-\nu}^{(n-\ell)}{}^*-\cpsi_{-\alpha}^{(\ell)}{}^*\cdpsi_{\nu-\alpha}^{(n-\ell)})\\
&\hspace{12mm}+\sum_{\ell=1}^{n-1}\Big(\sum_{\alpha\neq0,\nu+2}\cpsi_{\alpha}^{(\ell)}\cdpsi_{\nu+2-\alpha}^{(n-\ell)}-\sum_{\alpha\neq0,\nu-2}\cpsi_{-\alpha}^{(\ell)}{}^*\cdpsi_{\alpha+2-\nu}^{(n-\ell)}{}^*\Big)\bigg]\,,
\end{split}
\end{equation}
and similarly from Eq.~\eqref{eq:dpsinu} we get
\begin{equation}
\cdpsi_\nu^{(n)} = -\frac{\cdpsi_\nu^{(n-1)'}}{i\nu} + A_{\nu}^{(n)} + B_\nu^{(n)} + C_\nu^{(n)} + D_\nu^{(n)} \, ,
\end{equation}
where
\begin{equation}
\begin{split}
A_\nu^{(n)} & \equiv \\
 &-\frac{3}{2i\nu}\bigg[\cH_s\cdpsi^{(n-1)}_\nu+\cH_\nu^{(n-1)}\cdpsi_s+\sum_{\ell=1}^{n-1}\sum_{\alpha\neq0,\nu}\cH_\alpha^{(\ell)}\cdpsi_{\nu-\alpha}^{(n-\ell)}\bigg]\\
&+\frac{3}{2i\nu}\bigg[\cH_s\cdpsi_s\delta_{\nu,2}\delta_{n,2}+(1-\delta_{\nu,2})(\cH_s\cdpsi^{(n-1)}_{2-\nu}{}^*+\cH_{\nu-2}^{(n-1)}\cdpsi_s^*)+\sum_{\ell=1}^{n-1}\sum_{\alpha\neq0,\nu-2}\cH_\alpha^{(\ell)}\cdpsi_{\alpha+2-\nu}^{(n-\ell)}{}^*\bigg]\\
&+\frac{1}{2\nu}\bigg[\frac{\cnabla^2\cdpsi_s^*}{a_s^2}\delta_{\nu,2}\delta_{n,2}+\frac{\cnabla^2\cdpsi_\nu^{(n-1)}}{a_s^2}+(1-\delta_{\nu,2})(\frac{\cnabla^2\cdpsi_{2-\nu}^{(n-1)}{}^*}{a_s^2}+r_{\nu-2}^{(n-2)}\cnabla^2\cdpsi_s^*)\\
&\hspace{10mm}+r_\nu^{(n-2)}\cnabla^2\cdpsi_s+\sum_{\ell=1}^{n-2}\Big(\sum_{\alpha\neq0,\nu}r_\alpha^{(\ell)}\cnabla^2\cdpsi_{\nu-\alpha}^{(n-\ell-1)}+\sum_{\alpha\neq0,\nu-2}r_\alpha^{(\ell)}\cnabla^2\cdpsi_{\alpha+2-\nu}^{(n-\ell-1)}{}^*\Big)\bigg]\\
&-\frac{1}{\nu}\bigg[\Phi_s\cpsi^{(n-1)}_\nu+\Phi_\nu^{(n-1)}\cpsi_s+\Phi_s\cpsi_s\delta_{\nu,2}\delta_{n,2}+(1-\delta_{\nu,2})(\Phi_s\cpsi^{(n-1)}_{2-\nu}{}^*+\Phi_{\nu-2}^{(n-1)}\cpsi_s^*)\\
&\hspace{10mm}+\sum_{\ell=1}^{n-1}\Big(\sum_{\alpha\neq0,\nu}\Phi_\alpha^{(\ell)}\cpsi_{\nu-\alpha}^{(n-\ell)}+\sum_{\alpha\neq0,\nu-2}\Phi_\alpha^{(\ell)}\cpsi_{\alpha+2-\nu}^{(n-\ell)}{}^*\Big)\bigg] \,,
\end{split}
\label{Anun}
\end{equation}
\\
\\
\begin{equation}
\begin{split}
B_\nu^{(n)} &\equiv  \\ 
&\frac{-1}{2\nu}\bigg[\cpsi_s^2\cdpsi_{-\nu}^{(n-2)}{}^*+2\cpsi_s\cdpsi_s^*\cpsi_\nu^{(n-2)}+\cpsi_s^2\cdpsi_s\delta_{\nu,-2}\delta_{n,3}+\cpsi_s^*{}^2\cdpsi_s\delta_{\nu,2}\delta_{n,3}+\cpsi_s^*{}^2\cdpsi_s^*\delta_{\nu,4}\delta_{n,3}\\
&\hspace{10mm}+(1-\delta_{\nu,2})(\cpsi_s^*{}^2\cdpsi_{\nu-2}^{(n-2)}+2\cpsi_s^*\cdpsi_s\cpsi_{2-\nu}^{(n-2)}{}^*)+(1-\delta_{\nu,4})(\cpsi_s^*{}^2\cdpsi_{4-\nu}^{(n-2)}{}^*+2\cpsi_s^*\cdpsi_s\cpsi_{4-\nu}^{(n-2)}{}^*)\\
&\hspace{10mm}+(1-\delta_{\nu,-2})(\cpsi_s^2\cdpsi_{\nu+2}^{(n-2)}+2\cpsi_s\cdpsi_s\cpsi_{\nu+2}^{(n-2)})\\
&\hspace{10mm}+\sum_{\ell,\jmath=1}^{n-2}{'}\Big(\sum_{\alpha,\beta\neq0,\nu;\alpha+\beta\neq\nu}\cpsi_\alpha^{(\ell)}\cpsi_\beta^{(\jmath)}\cdpsi_{\alpha+\beta-\nu}^{(n-\ell-\jmath-1)}{}^*+\sum_{\alpha,\beta\neq0,\nu+2;\alpha+\beta\neq\nu+2}\cpsi_\alpha^{(\ell)}\cpsi_\beta^{(\jmath)}\cdpsi_{2+\nu-\alpha-\beta}^{(n-\ell-\jmath-1)}\Big)\\
&\hspace{10mm}+\sum_{\ell,\jmath=1}^{n-2}{'}\Big(\sum_{\alpha,\beta\neq0,\nu-4;\alpha+\beta\neq\nu-4}\cpsi_{-\alpha}^{(\ell)}{}^*\cpsi_{-\beta}^{(\jmath)}{}^*\cdpsi_{\alpha+\beta+4-\nu}^{(n-\ell-\jmath-1)}+\sum_{\alpha,\beta\neq0,\nu-2;\alpha+\beta\neq\nu-2}\cpsi_{-\alpha}^{(\ell)}{}^*\cpsi_{-\beta}^{(\jmath)}{}^*\cdpsi_{\nu-2-\alpha-\beta}^{(n-\ell-\jmath-1)}\Big)\\
&\hspace{10mm}+\sum_{\ell=1}^{n-2}\sum_{\alpha\neq0,\nu}(2\cpsi_s\cpsi_\alpha^{(\ell)}\cdpsi_{\alpha-\nu}^{(n-\ell-1)}{}^*+\cdpsi_s^*\cpsi_\alpha^{(\ell)}\cpsi_{\nu-\alpha}^{(n-\ell-1)})\\
&\hspace{10mm}+\sum_{\ell=1}^{n-2}\sum_{\alpha\neq0,\nu+2}(2\cpsi_s\cpsi_\alpha^{(\ell)}\cdpsi_{\nu+2-\alpha}^{(n-\ell-1)}{}^*+\cdpsi_s\cpsi_\alpha^{(\ell)}\cpsi_{\nu+2-\alpha}^{(n-\ell-1)})\\
&\hspace{10mm}+\sum_{\ell=1}^{n-2}\sum_{\alpha\neq0,\nu-2}(\cdpsi_s\cpsi_{-\alpha}^{(\ell)}{}^*\cpsi_{2+\alpha-\nu}^{(n-\ell-1)}{}^*+2\cpsi_s^*\cpsi_{-\alpha}^{(\ell)}{}^*\cdpsi_{\nu-2-\alpha}^{(n-\ell-1)})\\
&\hspace{10mm}+\sum_{\ell=1}^{n-2}\sum_{\alpha\neq0,\nu-4}(2\cpsi_s^*\cpsi_{-\alpha}^{(\ell)}{}^*\cdpsi_{4+\alpha-\nu}^{(n-\ell-1)}{}^*+\cdpsi_s^*\cpsi_{-\alpha}^{(\ell)}{}^*\cpsi_{4+\alpha-\nu}^{(n-\ell-1)}{}^*)\bigg]\,.
\end{split}
\label{Dnun}
\end{equation}
\begin{equation}
\begin{split}
C_\nu^{(n)} &\equiv \\
&\frac{2}{i\nu}\bigg[\cH_s\cpsi_s^*\Phi_s\delta_{\nu,2}\delta_{n,3}+(1-\delta_{\nu,2})(\cH_s\Phi_s\cpsi_{2-\nu}^{(n-2)}{}^*+\cH_s\cpsi_s^*\Phi_{\nu-2}^{(n-2)}+\Phi_s\cpsi_s^*\cH_{\nu-2}^{(n-2)})\\
&\hspace{7mm}-\cH_s\cpsi_s\Phi_\nu^{(n-2)}-\cH_s\Phi_s\cpsi_\nu^{(n-2)}-\Phi_s\cpsi_s\cH_\nu^{(n-2)}\\&\hspace{7mm}-\sum_{\ell=1}^{n-2}\sum_{\alpha\neq0,\nu}(\cH_s\Phi_{\alpha}^{(\ell)}\cpsi_{\nu-\alpha}^{(n-\ell-1)}+\Phi_s\cH_{\alpha}^{(\ell)}\cpsi_{\nu-\alpha}^{(n-\ell-1)}+\cpsi_s\cH_{\alpha}^{(\ell)}\Phi_{\nu-\alpha}^{(n-\ell-1)})\\
&\hspace{7mm}+\sum_{\ell=1}^{n-2}\sum_{\alpha\neq0,\nu-2}(\cH_s\Phi_{\alpha}^{(\ell)}\cpsi_{2+\alpha-\nu}^{(n-\ell-1)}{}^*+\Phi_s\cH_\alpha^{(\ell)}\cpsi_{2+\alpha-\nu}^{(n-\ell-1)}{}^*+\cpsi_s^*\cH_\alpha^{(\ell)}\Phi_{\nu-2-\alpha}^{(n-\ell-1)})\\
&\hspace{7mm}+\sum_{\ell,\jmath=1}^{n-2}{'}\Big(\sum_{\alpha,\beta\neq0,\nu-2;\alpha+\beta\neq\nu-2}\cH_\alpha\Phi_\beta\cpsi_{2+\alpha+\beta-\nu}{}^*-\sum_{\alpha,\beta\neq0,\nu;\alpha+\beta\neq\nu}\cH_\alpha^{(\ell)}\Phi_\beta^{(\jmath)}\cpsi_{\nu-\alpha-\beta}^{(n-\ell-\jmath-1)}\Big)\bigg]\, ,
\end{split}
\label{Cnun}
\end{equation}
and
\begin{equation}
\begin{split}
D_\nu^{(n)} &\equiv \\
&\frac{3}{\nu}\bigg[\cH_s^2\cdpsi_s^*\delta_{\nu,2}\delta_{n,3}+(1-\delta_{\nu,2})(\cH_s^2\cdpsi_{2-\nu}^{(n-2)}{}^*+2\cH_s\cdpsi_s^*\cH_{2-\nu}^{(n-2)})+\cH_s^2\cdpsi_\nu^{(n-2)}+2\cH_s\cdpsi_s\cH_\nu^{(n-2)}\\
&\hspace{7mm}+\sum_{\ell,\jmath=1}^{n-2}{'}\Big(\sum_{\alpha,\beta\neq0,\nu;\alpha+\beta\neq\nu}\cH_\alpha^{(\ell)}\cH_\beta^{(\jmath)}\cdpsi_{\nu-\alpha-\beta}^{(n-\ell-\jmath-1)}
+\sum_{\alpha,\beta\neq0,\nu-2;\alpha+\beta\neq\nu-2}\cH_\alpha^{(\ell)}\cH_\beta^{(\jmath)}\cdpsi_{2+\alpha+\beta-\nu}^{(n-\ell-\jmath-1)}{}^*\Big)\\
&\hspace{7mm}+\sum_{\ell=1}^{n-2}\sum_{\alpha\neq0,\nu}(2\cH_s\cH_\alpha^{(\ell)}\cdpsi_{\nu-\alpha}^{(n-\ell-1)}+\cdpsi_s\cH_\alpha^{(\ell)}\cH_{\nu-\alpha}^{(n-\ell-1)})\\
&\hspace{7mm}+\sum_{\ell=1}^{n-2}\sum_{\alpha\neq0,\nu-2}(2\cH_s\cH_\alpha^{(\ell)}\cdpsi_{2+\alpha-\nu}^{(n-\ell-1)}{}^*+\cdpsi_s^*\cH_\alpha^{(\ell)}\cH_{\nu-2-\alpha}^{(n-\ell-1)})\bigg] \, .
\end{split}
\label{Bnun}
\end{equation}
In the above expressions, a prime over the summation means $\ell+\jmath$ must be less than the upper limit. Most of the above terms are zero at leading order, but at higher orders more and more terms contribute. For instance it can be shown that for $n=3$ we have 
\eq{
\spl{
\cdpsi_\nu^{(3)}=&\delta_{\nu,2}\left[\left(-\frac{3}{32}\cpsi_s^*{}^2+\frac{1}{16}|\cpsi_s|^2+\frac{\cnabla^4}{8a_s^4}+\frac{5\cH_s\cnabla^2}{8ia_s^2}\right)\cdpsi_s^*-\frac{\cpsi_s^*{}^2}{8}\cdpsi_s\right]\\
&+\delta_{\nu,-2}\frac{13}{32}\cpsi_s^2\cdpsi_s-\delta_{\nu,4}\frac{13}{64}\cpsi_s^*{}^2\cdpsi_s^*\,,
}
}
and
\eq{
\Phi_\nu^{(3)}=\delta_{\nu,2}\left[\cpsi_s^*\left(\frac{\cnabla^2}{16a_s^2}+\frac{\cH_s}{8i}\right)\cdpsi_s^*-\frac{1}{16}\cpsi^*{}^2\Phi_s\right]+\delta_{\nu,-2}\left[\cpsi_s\left(\frac{\cnabla^2}{16a_s^2}-\frac{\cH_s}{8i}\right)\cdpsi_s-\frac{1}{16}\cpsi^2\Phi_s\right]\,.
}
This procedure can be continued straightforwardly to obtain higher orders. These terms can then be used to get an effective equation for the slow modes of different variables as done in Sec.~\ref{sec:EFT}. After lengthy algebra we obtain the effective equation for $\dpsi_s$, up to this order, as follows
\eq{
\spl{
\label{eq:eftsch2}
&i\dot{\dpsi_s}+i\frac{3}{2}H_s\dpsi_s+\frac{\nabla^2\dpsi_s}{2ma_s^2}-m \bar{\psi}_s \Phi_s\\
&+\frac{9}{8}\frac{|\bar{\psi}_s|^2}{\Mpl^2}\dpsi_s - \frac{7}{16} \frac{\bar{\psi}_s^2}{\Mpl^2} \dpsi_s^* + \frac{\nabla^4\dpsi_s}{8m^3a_s^4}+ 2iH_s\bar{\psi}_s\Phi_s\\
&+\left(\frac{3i|\bar{\psi}_s|^2H_s}{16m\Mpl^2} + \frac{17| \bar{\psi}_s|^2 \nabla^2}{32m^2\Mpl^2a_s^2} - \frac{H_s\nabla^4}{8im^4a_s^4} + \frac{\nabla^6}{16m^5a_s^6} \right) \dpsi_s + \left(\frac{57i\bar{\psi}_s^2H_s}{32m\Mpl^2} - \frac{7\bar{\psi}_s^2 \nabla^2}{32m^2\Mpl^2a_s^2}\right) \dpsi_s^*\\
&+\frac{7|\bar{\psi}_s|^2 \bar{\psi}_s}{8\Mpl^2}\Phi_s=0\,.
}
}
Likewise, for $\Phi_s$ we have
\eq{
\spl{
\label{eq:eftphidot2}
&\dot{\Phi}_s + H_s \Phi_s + \frac{i}{4\Mpl^2}(\bar{\psi}_s \dpsi_s^* - \bar{\psi}_s^* \dpsi_s)\\
& + \frac{3H_s}{8m\Mpl^2}(\bar{\psi}_s \dpsi_s^* + \bar{\psi}_s^* \dpsi_s) + \frac{i\nabla^2}{16m^2\Mpl^2a_s^2} (\bar{\psi}_s \dpsi_s^* - \bar{\psi}_s^*\dpsi_s)\\
&+\frac{i|\bar{\psi}_s|^2}{16m\Mpl^4} (\bar{\psi}_s \dpsi_s^* - \bar{\psi}_s^* \dpsi_s) + \frac{13H_s\nabla^2}{64m^3\Mpl^2a_s^2} (\bar{\psi}_s \dpsi_s^* + \bar{\psi}_s^* \dpsi_s) + \frac{i\nabla^4}{32m^4\Mpl^2a_s^4}(\bar{\psi}_s \dpsi_s^*- \bar{\psi}_s^* \dpsi_s)\\
&-\frac{3H_s|\bar{\psi}_s|^2}{16m\Mpl^2}\Phi_s=0\,.
}
}
The effective Poisson equation takes form
\eq{
\spl{
\label{eq:eftpoisson22}
\frac{\nabla^2\Phi_s}{a_s^2} = &\frac{m}{2\Mpl^2}(\bar{\psi}^*_s \dpsi_s + \bar{\psi}_s\dpsi^*_s)\\
&+\frac{3iH_s}{4\Mpl^2}(\bar{\psi}^*_s \dpsi_s - \bar{\psi}_s \dpsi^*_s) - \frac{3}{2}H_s^2\Phi_s\\
&+\frac{11|\bar{\psi}_s|^2}{32\Mpl^4} (\bar{\psi}_s \dpsi_s^* + \bar{\psi}_s^* \dpsi_s) + \frac{3iH_s\nabla^2}{32m^2\Mpl^2a_s^2}(\bar{\psi}_s \dpsi_s^*-\bar{\psi}_s^*\dpsi_s)\,.
}
}

\section{Viscous fluid}\label{app:vis}
In this appendix we review the equations for the evolution for an imperfect fluid, which are discussed in various references; see, e.g., Refs.~\cite{Landau:1959,Weinberg:1972kfs,Weinberg:2008zzc,Romatschke:2009im}. When describing an imperfect fluid, one typically characterizes the deviation from a perfect fluid by $\Delta T_{\mu\nu}$:
\eq{
T_{\mu\nu}=pg_{\mu\nu}+(p+\rho)u_\mu u_\nu+\Delta T_{\mu\nu}\,,
}    
where $p$, $\rho$ and $u^\mu$ are the pressure, density, and 4-velocity of the fluid. Here we adopt the convention of Refs.~\cite{Landau:1959,Weinberg:2008zzc} that $u^i$ is the velocity of energy transport. In general, for first-order hydrodynamics, one may consider effects of shear and bulk viscosity as well as heat transfer. However, for the purpose of our effective fluid description, up to the working order of Sec.~\ref{sec:fluid} we only need to consider the bulk viscosity to fully describe the system. At higher order, on the hand, we expect the effective shear viscosity and the effective heat transfer to appear (since these contributions are consistent with the symmetries of the low-energy effective theory).  In this simplified case $\Delta T_{\mu\nu}$ takes the form
\eq{
\Delta T_{\mu\nu}=-\zeta(g_{\mu\nu}+u_\mu u_\nu)u^\kappa{}_{;\kappa}\,,
} 
where $\zeta$ is the coefficient of bulk viscosity (which can be both time and position dependent the latter of which will be ignored here); and a semicolon denotes covariant differentiation. Note that the form of $\Delta T_{\mu\nu}$ suggests modification of the pressure $p$ with bulk viscosity pressure $\Pi_{bv}$ of the form: $p\to p+\Pi_{bv}$ where $\Pi_{bv}=-\zeta u^\kappa{}_{;\kappa}$ is proportional to the divergence of the velocity of the fluid. For the background equations this causes a trivial modification in the continuity equation of the form
\eq{
\label{eq:enbk}
\dot{\rho}+3H(\rho+p_{\text{eff}})=0\,,
}
where we have defined $p_{\text{eff}}=p-3H\zeta$. As a result, the effect of bulk viscosity at the background level is degenerate with the effect of pressure. In the language of our effective fluid description for the background evolution, the effective pressure in Eq.~\eqref{peff} must include the bulk viscosity term, i.e. it is the expression for $p_{\text{eff}}$. 

On the other hand, for fluctuations of the fluid, the bulk viscosity has nontrivial effects in the equations of motion, and the aforementioned degeneracy beteween pressure and the bulk viscosity will be broken. The energy conservation for fluctuations results in
\eq{
\dot{\dro}+3H(\dro+\dP)+(\rho+p_{\text{eff}})\frac{\nabla^2\du}{a^2}-3(\rho+p_{\text{eff}})\dot{\Phi}+9H\zeta \left( \dot{\Phi}+H\Phi-\frac{\nabla^2\du}{3a^2} \right)=0\,,
} 
where $\dro$ and $\dP$ are fluctuations of density and pressure, respectively, and $\du$ is the fluctuation in the velocity potential. Note that for fluctuations of the bulk viscosity pressure we set $\delta\Pi_{bv}=-\zeta\delta(u^\kappa{}_{;\kappa})$, i.e., we will not consider fluctuations of $\zeta$. We will see that this is sufficient to fully describe the system to the working order. Other equations for the fluctuations read
\ba 
\dP+\dot{p}_{\text{eff}}\, \du +(\rho+p_{\text{eff}})(\dot{\du}+\Phi)-\frac{3\zeta(\rho+p_{\text{eff}})}{2\Mpl^2}\du-\zeta\frac{\nabla^2\du}{a^2}=0 \, , 
\\
\dot{\Phi}+H\Phi+\frac{(\rho+p_{\text{eff}})}{2\Mpl^2}\du =0\,,
\\
\frac{\nabla^2\Phi}{a^2}=\frac{1}{2\Mpl^2}[\dro-3H(\rho+p_{\text{eff}})\du]\,.
\ea 
We must add to these equations a relation between pressure and the density fluctuations. This is usually done in the comoving gauge, defined by the condition $\du_c=0$, where we define
\eq{
\dP_c=\cs \, \dro_c\, .
\label{eq:dpdrhoc}
}
Here the subscript `$c$' denotes that Eq.~(\ref{eq:dpdrhoc}) holds in the comoving gauge, and $c_s$ is the sound speed \cite{Romano:2015vxz}. By a gauge transformation, we can find a similar relation in a general gauge as follows:
\eq{
\dP=\cs \left[\dro-3H(\rho+p_{\text{eff}})\du \right]-\left[\dot{p}_{\text{eff}} \, -\frac{3\zeta(\rho+p_{\text{eff}})}{2\Mpl^2}+3H\dot{\zeta}\right]\du\,,
}
where we have used Eq.~(\ref{eq:enbk}) for the background energy conservation. Defining the comoving overdensity by
\eq{
\delta\equiv\frac{1}{\rho}\left[ \dro-3H(\rho+p_{\text{eff}})\du \right]\,,
}
and using other equations (and their time derivatives) to remove all variables other than $\delta$, we obtain the following second-order differential equation for $\delta$:
\eq{
\label{eq:ddot_fluid}
\ddot{\delta}+2\gamma H\dot{\delta}-\left[\cs-\zeta\frac{3H p_{\text{eff}}}{\rho(\rho+p_{\text{eff}})}\right]\frac{\nabla^2\delta}{a^2}-\frac{\zeta}{(\rho+p_{\text{eff}})}\frac{\nabla^2\dot{\delta}}{a^2}=\frac{\vartheta\rho}{2\Mpl^2}\delta\,,
}
in which we have defined
\eqa{
\gamma&=1-\frac{3p_{\text{eff}}}{\rho}+\frac{3\dot{p}_{\text{eff}}}{2\dot \rho}-\frac{3\dot{\zeta}}{2(\rho+p_{\text{eff}})}\\
\vartheta&=1+\frac{8p_{\text{eff}}}{\rho}-\frac{3p_{\text{eff}}^2}{\rho^2}-\frac{6\dot{p}_{\text{eff}}}{\dot \rho}-\frac{6p_{\text{eff}}\dot{\zeta}}{\rho(\rho+p_{\text{eff}})}\,.
}
For the special case of $p_{\text{eff}}=0$ we have a simpler form
\eq{
\label{eq:ddot_fluidp0}
\ddot{\delta}+2H\left[1-\frac{3\dot{\zeta}}{2\rho}\right]\dot{\delta}-\cs\frac{\nabla^2\delta}{a^2}-\frac{\zeta}{\rho}\frac{\nabla^2\dot{\delta}}{a^2}=\frac{\rho}{2\Mpl^2}\delta\,.
}
This is the equation that is of our interest in Sec.~\ref{sec:fluid} since there we are working to ${\cal O}(\epsilon)$ for the field fluctuations, and to that order, $p_{\text{eff}}$ vanishes; its nonzero value only appears at ${\cal O}(\epsilon^2)$ and higher, which will be ignored. However, when we consider self-interaction in Appendix \ref{app:interaction} the effective pressure is nonzero at leading order and we must use Eq.~\eqref{eq:ddot_fluid}. Other fluid fluctuations can be written in terms of $\delta$ as 
\eqa{
\label{eq:poisdelta}
&\frac{\nabla^2\du}{a^2}=\frac{-\rho\dot{\delta}+3Hp_{\text{eff}}\delta}{\rho+p_{\text{eff}}} \, , \qquad  \qquad  
\frac{\nabla^2\Phi}{a^2}=\frac{\rho}{2\Mpl^2}\, \delta \\
&\dro=\rho\delta+3H(\rho+p_{\text{eff}})\du\, , \qquad \quad
\dP=\cs\rho\delta-\left[\dot{p}_{\text{eff}} \, -\frac{3\zeta(\rho+p_{\text{eff}})}{2\Mpl^2}+3H\dot{\zeta}\right] \, \du\,.
}
Before concluding this section we report similar equations in comoving gauge, i.e. $\du_c=0$ (which will be used in Appendix~\ref{app:gauge}). We write the line element in this gauge by 
\eq{
	\dd{s}^2=-(1+2N)\dd{t}^2+2a(t)\p_i\sigma\dd{t}\dd{x^i}+a(t)^2(1+2\mathcal{R})\delta_{ij}\dd{x^i}\dd{x^j}\, .
}
The equations are then found to take the following form: 
\eq{
\label{eq:comov2}
\dot{\mathcal{R}}=H \, N\,,\qquad\dot{\sigma}+2H\, \sigma+\frac{1}{a}(N+\mathcal{R})=0\,,\qquad\frac{\nabla^2\mathcal{R}}{a^2}+\frac{H\nabla^2\sigma}{a^2}+\frac{\dro_c}{2\Mpl^2}=0\,,
}
and, for the continuity and Euler equations we have
\eq{
\label{eq:comov3}
\dot{\dro}_c+3H\dro_c-(\rho+p_{\text{eff}})\frac{\nabla\sigma}{a}=0\,,\qquad\dP_c+(\rho+p_{\text{eff}})N+\zeta\frac{\nabla\sigma}{a}=0\,.
}

\section{EFT for the self-interacting field}
\label{app:interaction}

In this appendix we consider a self-interaction term in the potential,
\eq{
	V(\phi)=\frac{1}{2}m^2\phi^2+\frac{1}{4!}\lambda\phi^4\,.
}
Such self-interactions, even with a very weak coupling $\lambda$, might play an important role in the context of cold dark matter scenarios \cite{Guth:2014hsa,Fan:2016rda,Tulin:2017ara,Namjoo:2017nia}. We follow the same procedure as in Sec.~\ref{sec:EFT} to construct a low-energy EFT. We perform the field redefinition introduced in Sec.~\ref{sec:redef}, expand the new field $\psi$ as an infinite series of different modes, and find the equation of motion for each mode. Then we identify appropriate small parameters, perform a perturbative expansion of the nonzero modes, and incorporate the backreaction from (rapidly oscillating) nonzero modes into the effective equations of motion for the slow modes. To identify appropriate small parameters, we consider the equation of motion in terms of $\phi$ as discussed in the introduction. Since we are interested in the regime of oscillating solutions with dominant frequency $m$, we must have
\eq{
	\epsilon_\lambda\sim\lambda\frac{\vert \psi^2\vert}{m^3}\ll1\,,
} 
where we have expressed what we had in Eq.~\eqref{eq:eps} in terms of $\psi$. This new parameter, together with $\epsilon_t$, $\epsilon_{\scriptscriptstyle H}$, and $\epsilon_k$, forms the set of small parameters for our problem. As discussed in Sec.~\ref{sec:EFT}, for fluctuations, we also consider 
the small parameter $\epsilon_g$, to indicate that we are working only to linear order in spatially varying fluctuations and make the power counting more straightforward.
Here we present the results of our analysis up to $\order{\epsilon^2}$ for background and $\order{\epsilon}$ for fluctuations.  For background variables we have

\eqa{
	\label{eq:eftbckslf}
	\nonumber
	&i\dot{\bar{\psi}}_s + i\frac{3}{2}H_s \bar{\psi}_s - \lambda\frac{|\bar{\psi}_s|^2 \bar{\psi}_s}{8m^2} + \frac{9}{16} \frac{|\bar{\psi}_s|^2}{\Mpl^2} \bar{\psi}_s + \lambda^2\frac{17|\bar{\psi}_s|^4 \bar{\psi}_s}{768m^5} + \frac{9i}{32\Mpl^2}H_s|\bar{\psi}_s|^2 \bar{\psi}_s
	\\
	&\quad \, \, \, +\frac{3\lambda|\bar{\psi}_s|^4 \bar{\psi}_s}{64m^3\Mpl^2} + \frac{27i\lambda^2H_s|\bar{\psi}_s|^4 \bar{\psi}_s}{1024m^6} - \frac{23\lambda^3|\bar{\psi}_s|^6 \bar{\psi}_s}{73728m^8}
	+\order{\epsilon^3}H_s \bar{\psi}_s=0\,,\\\nonumber \\
	&3\Mpl^2H_s^2=m|\bar{\psi}_s|^2 + \lambda\dfrac{|\bar{\psi}_s|^4}{16m^2} + \dfrac{3|\bar{\psi}_s|^4}{32\Mpl^2} - \frac{95\lambda^2|\bar{\psi}_s|^6}{9216m^5} + \order{\epsilon^3}m|\bar{\psi}_s|^2\,,
}
as the effective Schr\"odinger and Friedmann equations, respectively. Note that the  term proportional to $\lambda^2$ on the first line in Eq.~\eqref{eq:eftbckslf} has also been obtained in Ref.~\cite{Namjoo:2017nia} in Minkowski spacetime. Likewise, for fluctuations we have 
\eq{
	\spl{
		&i\dot{\dpsi}_s + \frac{3}{2}i H_s \dpsi_s + \frac{\nabla^2\dpsi_s}{2ma_s^2} - m\bar{\psi}_s \Phi_s - \frac{\lambda\bar{\psi}_s^2 \dpsi_s^*}{8m^2} - \frac{\lambda|\bar{\psi}_s|^2 \dpsi_s}{4m^2} + \frac{9|\bar{\psi}_s|^2}{8\Mpl^2} \dpsi_s - \frac{7\bar{\psi}_s^2}{8\Mpl^2} \dpsi_s^*\\
		&+\frac{\nabla^4\dpsi_s}{8m^3a_s^4} + 2iH_s \bar{\psi}_s \Phi_s - \frac{\lambda|\bar{\psi}_s|^2 \nabla^2 \dpsi_s}{8m^4a_s^2} - \frac{\lambda \bar{\psi}_s^2 \nabla^2 \dpsi_s^*}{16m^4a_s^2} + \frac{17\lambda^2|\bar{\psi}_s|^4 \dpsi_s}{256m^4} + \frac{17\lambda^2|\bar{\psi}_s|^2 \bar{\psi}_s^2\dpsi_s^*}{384m^4}
		\\ 
		& +\order{\epsilon^2} H_s \delta \bar{\psi}_s 
		=0\,,
		\label{eq:fluctslambda}
	}
}
as the effective equation for $\dpsi_s$. The equation for $\Phi_s$ is the same as the non-interacting case in Eq.~\eqref{eq:eftphidot}. The effective Poisson equation takes the form
\eq{
	\spl{
		\label{eq:eftpoisson2}
		\frac{\nabla^2\Phi_s}{a_s^2}= &\frac{m}{2M_P^2}(\bar{\psi}^*_s \dpsi_s + \bar{\psi}_s \dpsi^*_s)\\
		&+\frac{3i}{4M_P^2}H_s (\bar{\psi}^*_s \dpsi_s - \bar{\psi}_s \dpsi^*_s) - \frac{3}{2}H_s^2 \Phi_s + \frac{\lambda|\bar{\psi}_s|^2}{16m^2\Mpl^2} (\bar{\psi}^*_s \dpsi_s + \bar{\psi}_s \dpsi^*_s)
\\	
&+\order{\epsilon^2} H_s \Phi_s =0 \, .
	}
}
We can find the effective fluid density and pressure by the similar procedure outlined in Sec.~\ref{sec:fluid}, which results in
\ba
\nonumber
\rho_{\text{eff}} = m|\bar{\psi}_s|^2 + \lambda\frac{|\bar{\psi}_s|^4}{16m^2} + \frac{3|\bar{\psi}_s|^4}{32\Mpl^2} - \frac{95\lambda^2|\bar{\psi}_s|^6}{9216m^5}
\, , \qquad 
	p_{\text{eff}}=\lambda\frac{|\bar{\psi}_s|^4}{16m^2} + \frac{3|\bar{\psi}_s|^4}{32\Mpl^2} - \frac{95\lambda^2|\bar{\psi}_s|^6}{4608m^5}\,.
\\
\ea
The effective comoving overdensity can be read from Eq.~\eqref{eq:eftpoisson2}:
\eq{
	\label{eq:deltaeffdef2}
	\delta_{\text{eff}} = \left(\frac{\dpsi_s}{\bar{\psi}_s} + \frac{\dpsi_s^*}{\bar{\psi}_s^*} \right) + i\frac{3H_s}{2m} \left( \frac{\dpsi_s}{\bar{\psi}_s} - \frac{\dpsi_s^*}{\bar{\psi}_s^*} \right) + \frac{\lambda|\bar{\psi}_s|^2}{16m^3} \left(\frac{\dpsi_s}{\bar{\psi}_s} + \frac{\dpsi_s^*}{\bar{\psi}_s^*}\right)-\Phi_s\,.
}
We then obtain a second-order differential equation for $\delta_{\text{eff}}$ from which one can read the effective sound speed and the effective viscosity. Comparing with Eq.~\eqref{eq:ddot_fluid} we obtain 
\eq{
	\label{eq:eftcslambda}
	\spl{
		c^2_{\text{eff}} = &\frac{k^2}{4m^2a_s^2} + \frac{\lambda|\bar{\psi}_s|^2}{8m^3}\\
		&-\frac{k^4}{8m^4a_s^4} + \frac{5|\bar{\psi}_s|^2}{16m\Mpl^2} - \frac{5k^2\lambda|\bar{\psi}_s|^2}{32m^2a^2} - \frac{23\lambda^2|\bar{\psi}_s|^4}{384m^6}\,,
	}
}
where the leading term proportional to $\lambda$ is consistent with the results of Refs.~\cite{Fan:2016rda,Brax:2019fzb}. And, finally, the bulk viscosity coefficient reads
\eq{
	\label{eq:eftzetalambda}
	\zeta_{\text{eff}}=-\frac{H_s}{2m^2}(\rho_{\text{eff}}+p_{\text{eff}})\,.
}

\section{Analyses in a multicomponent universe}
\label{app:multi-component}
Until now the scalar field was considered to be the only matter component in the universe. However, in a realistic situation, other components are present. As a step forward,  we now consider a perfect fluid component (with a constant equation of state) along with the scalar field. This component can be another dark matter species in a matter-dominated universe or radiation as is the case, e.g., around the time of recombination. At the background level by energy conservation we have
\eq{
	\label{eq:rhos}
	\dot{\rho}+3H(\rho+p)=0\, .
}
Note that we are assuming that the scalar field and the new fluid component are interacting only gravitationally. The Schr\"odinger equation, Eq.~\eqref{eq:bckg1}, is the same as before, and the Friedmann equation now reads
\eq{
	3\Mpl^2H^2=m|\psi|^2+\rho\, .
}
We are not making any assumption about the relative fraction of the energy densities. However, we expect that by decreasing the share of the scalar field, the backreaction effects of rapid oscillations will be suppressed. We, of course, still need to assume that $H\ll m$, such that the scalar field is oscillating, so our formalism is applicable. We present the results considering terms up to order $n=3$ (inclusive) and hence up to $\order{\epsilon^2}$ in the effective equations for the slow modes:
\eq{
	i\dot{\bar{\psi}}_s + i\frac{3}{2} H_s \bar{\psi}_s + \frac{3}{8m\Mpl^2} \left( \frac{3}{2}m|\bar{\psi}_s|^2 + i\rho_s\right) \bar{\psi}_s + \frac{9iH_s \bar{\psi}_s}{32m^2\Mpl^2} \left(m|\bar{\psi}_s|^2 + \rho_s + p_s\right) + \order{\epsilon^3}H_s \bar{\psi}_s=0\,,
}   
where $\rho_s$ and $p_s$ are the slow-mode parts of the density and pressure of the additional fluid. The effective Friedmann equation reads
\eq{
	\label{eq:fridother}
	3\Mpl^2H_s^2 = m|\bar{\psi}_s|^2 + \rho_s + \frac{3|\bar{\psi}_s|^2}{16m\Mpl^2} \left( \frac{1}{2}m| \bar{\psi}_s|^2 + \rho_s \right)+\order{\epsilon^3}m|\bar{\psi}_s|^2\,.
}
The effective equation of energy conservation for $\rho_s$, at this order, is the same as Eq.~\eqref{eq:rhos} and does not get corrections, i.e. $\dot{\rho}_s+3H_s(\rho_s+p_s)=0$. 

We now proceed to consider fluctuations. The fluctuations for a perfect fluid are characterized by $\dro$, $\dP$ and $\du$ corresponding to fluctuations in  energy density, pressure, and the velocity potential. The relevant, exact equations are 
\eqa{
	&i\dot{\dpsi}+[\mathfrak{D} \dpsi-\mathfrak{J} \bar{\psi}]
	+e^{2imt}[\mathfrak{D}^*\dpsi^*-\mathfrak{J}^* \bar{\psi}^*]=0 ,\\
	\label{eq:enc}
	&\dot{\dro}+3H(\dro+\dP)+(\rho+p)\left(\frac{\nabla^2\du}{a^2}-3\dot{\Phi}\right)=0\,,\\
	\label{eq:mmc}
	&\partial_t((\rho+p)\du)+(\rho+p)(3H\du+\Phi)+\dP=0\,,\\
	&\dot{\Phi} + H\Phi + \frac{i}{4\Mpl^2} \left( \bar{\psi}\dpsi^* - \bar{\psi}^*\dpsi + e^{-2imt} \bar{\psi}\dpsi - e^{2imt}\bar{\psi}^* \dpsi^* \right) + \frac{1}{2\Mpl^2}(\rho+p)\du=0\,,\\
	&2\Mpl^2\frac{\nabla^2\Phi}{a^2} = m( \bar{\psi} \dpsi^* + \bar{\psi}^* \dpsi) + \dro + 6\Mpl^2H\dot{\Phi} + \left(\frac{m}{2}(e^{-imt}\bar{\psi} + e^{imt} \bar{\psi}^*)^2 + 2\rho\right)\Phi\,,
} 
where we have defined 
\eq{
	\mathfrak{J} \equiv (m-2iH)\Phi-\frac{i(\rho+p)\du}{\Mpl^2}\,,
}
and $\mathfrak{D}$ is defined in Eq.~\eqref{eq:Df}. We also need the relation $\dP_c=\tilde c_s^2\dro_c$, in comoving gauge $\du_c=0$, where $\tilde c_s$ is the speed of sound for the new component (not to be confused with the sound speed deduced from the EFT for the scalar field), which can be written in Newtonian gauge via a gauge transformation. Again following the same set of steps as in Sec.~\ref{sec:EFT}, we obtain the effective equations for slow-mode variables $\dpsi_s$, $\Phi_s$, as well as $\dro_s$, $\dP_s$, and $\du_s$. Here we only work up to order $n=2$ (which implies that in the effective equations for the slow modes, we neglect terms at the order of $\epsilon^2$ and higher). At this order, equations of energy and momentum conservation for the fluid remain unchanged, i.e. Eqs.~\eqref{eq:enc} and \eqref{eq:mmc} hold with all variables replaced by their slow-mode counterparts. The equation for $\dpsi_s$ at this order becomes
\eq{
	\spl{
		&i\dot{\dpsi_s} + i\frac{3}{2} H_s \dpsi_s + \frac{\nabla^2\dpsi_s}{2ma_s^2} - m\bar{\psi}_s\Phi_s\\
		&+\frac{3}{8m\Mpl^2} (3m|\bar{\psi}_s|^2 + \rho_s) \dpsi_s - \frac{7}{16}\frac{\bar{\psi}_s^2}{\Mpl^2} \dpsi_s^* + \frac{\nabla^4\dpsi_s}{8m^3a_s^4} + \left( \frac{i(\rho_s+p_s)\du_s}{\Mpl^2} - \frac{\dro_s}{8m\Mpl^2} + 2iH_s\Phi_s\right)\bar{\psi}_s
		\\
		&  +\order{\epsilon^2} H_s \delta \bar{\psi}_s=0 \,  ,
	}
}
and for $\Phi_s$ with only a few changes
\eq{
	\spl{
		&\dot{\Phi}_s + H_s\Phi_s + \frac{i}{4\Mpl^2}(\bar{\psi}_s \dpsi_s^* - \bar{\psi}_s^*\dpsi_s)\\
		&+\frac{3H_s}{8m\Mpl^2}(\bar{\psi}_s \dpsi_s^* + \bar{\psi}_s^* \dpsi_s) + \frac{i}{8m\Mpl^2}\frac{\nabla^2}{2ma_s^2}(\bar{\psi}_s \dpsi_s^* - \bar{\psi}_s^* \dpsi_s) + \frac{1}{2\Mpl^2}(\rho_s+p_s)\du_s
		\\
		&  +\order{\epsilon^2} H_s \Phi_s =0\, ,
	}
}
and
\eq{
	\spl{
		\frac{\nabla^2\Phi_s}{a_s^2}
&	=\frac{1}{2\Mpl^2}(m \bar{\psi}^*_s \dpsi_s + m \bar{\psi}_s \dpsi^*_s + \dro_s)
		\\
	&	+\frac{3i}{4\Mpl^2}H_s( \bar{\psi}^*_s \dpsi_s - \bar{\psi}_s \dpsi^*_s ) - \frac{m|\bar{\psi}_s|^2}{2\Mpl^2} \Phi_s - \frac{3H_s(\rho_s+p_s)}{2\Mpl^2}\du_s
		 +\order{\epsilon^2} H_s \Phi_s \, .
	}
}
As for the effective fluid description, it is rather tricky to derive the corresponding sound speed and viscosity in the case of a multicomponent universe, as the definition of the density contrast for the individual components is ambiguous (partly because we are dealing with two systems that are not well separated and indeed coupled through gravity). We will leave this analyses for a future work and only present here the effective energy density and the effective pressure for the background evolution: 
\eq{
	\rho_{\rm eff}=m|\bar{\psi}_s|^2 + \frac{3|\bar{\psi}_s|^2}{16m\Mpl^2} \left( \frac{1}{2}m|\bar{\psi}_s|^2+\rho_s\right)\,,
}
and
\eq{
	p_{\rm eff}=\frac{9|\bar{\psi}_s|^4}{32\Mpl^2} + \frac{3|\bar{\psi}_s|^2}{8m\Mpl^2}(p_s+\rho_s)\, .
}

\section{Nonlocal field redefinition and the sound speed at arbitrary scale}\label{app:nonlocal}
In Ref.~\cite{Namjoo:2017nia} a nonlocal field redefinition is introduced, which yielded a dramatic simplification when deriving an EFT in Minkowski spacetime. In this appendix, we use the same field redefinition --- still in Minkowski spacetime --- to derive a sound speed for the fluctuations, which we expect to hold for arbitrary momentum. We show that this leads to a sound speed, consistent with Eq.~\eqref{eq:eftcs} in low-momentum limit, thereby confirming our results, including the coefficient of the subdominant term. Further, this sound speed converges to $c_s \simeq 1$ in the relativistic regime, as expected. We also add to the analysis of Ref.~\cite{Namjoo:2017nia} by {\it deriving} the nonlocal operator (rather than simply postulating it), by requiring the resulting theory to have certain properties. We will see that these requirements do not fix the field redefinition uniquely, but they do suggest that the redefinition proposed in Ref.~\cite{Namjoo:2017nia} is the simplest possibility. Taking some steps more generally to also include the case of an unperturbed FLRW background, we will see that a similar set of requirements \emph{fail} in an FLRW universe to give sufficiently simple relations, which justifies our approach in this paper of returning to a local field redefinition.   

We start from the Lagrangian in terms of $\phi$ and $\chi$ given in Eq.~\eqref{eq:phichilag} and consider a more general field redefinition to introduce the  $\psi$ field. Note that $\phi$ and $\chi$ are both real fields, therefore a general field redefinition may take the following form:
\eqa{
	\label{phi_N}
	\phi = \frac{1}{\sqrt{2m}}  \left[O_1\,\psi + O_1^*\, \psi^* \right], \qquad 
	\chi = -i\sqrt{\frac{m}{2}} \, \left[O_2\, \psi -O_2^*\, \psi^* \right]\,,
} 
in which $O_1$ and $O_2$ can be considered as arbitrary operators which can depend explicitly on time, space, time derivatives, and spatial derivatives. We restrict these general functions by the following considerations. First, we do not want to introduce any extra, non-physical degrees of freedom, so we assume that $O_1$ and $O_2$ are independent of the time-derivative operator. This avoids higher-order time derivatives in the resulting equations of motion. Second, we want both sides of the above equations to be 3-scalars so that the final Lagrangian is a 3-scalar. (In an FLRW universe, the time diffeomorphism is broken, so we do not require the Lagrangian to be a 4-scalar.) This means that, in the absence of any other degrees of freedom or preferred direction, the arbitrary functions $O_1$ and $O_2$ only depend on spatial derivatives via $\nabla_i \nabla^i$, where $\nabla_i$ is the covariant derivative on spatial hypersurfaces. Third, we require that in the zero-momentum limit we recover the traditional field transformation. By comparing Eqs.~\eqref{phi_N} and \eqref{eq:cantr1}, this yields
\eq{
	\label{eq:minlim}
	O_1(\nabla_i \nabla^i, {\bf x},t)\to e^{-imt} + \order{\nabla^2/m^2} \qquad \text{for} \quad \nabla^2/m^2 \to 0.
}
This ensures that we obtain the Schr\"odinger equation in the same limit.  
Finally, we want the $\psi$ field to be a 3-scalar, and to transform under spatial rotations in the standard way, which restricts the spatial dependence of $O_1$ and $O_2$. We simply assume that $O_1$ and $O_2$ retain no dependence on ${\bf x}$. With these requirements, the arbitrary functions $O_1$ and $O_2$ take the forms $O_1 = O_1 (\nabla_i \nabla^i, t)$ and $O_2 = O_2(\nabla_i \nabla^i, t)$, with the additional restriction of Eq.~\eqref{eq:minlim} for $O_1$. Note that with these restrictions, $O_1$ and $O_2$ commute in an unperturbed FLRW universe.

To further restrict the forms of $O_1$ and $O_2$, we consider the Lagrangian in terms of $\psi$ and $\psi^*$. Plugging the definitions of Eq.~\eqref{phi_N} into the Lagrangian of Eq.~\eqref{eq:phichilag}, assuming an unperturbed FLRW spacetime --- that is, neglecting backreaction of the field on the spacetime geometry --- and performing some integrations by parts we obtain
\eq{
	\label{eq:lagnnlocal}
	\mathcal{L}=\frac{ia^3}{2}\left(\dot{\psi}\Gamma_1\psi^*-\dot{\psi}^*\Gamma_1^*\psi\right)+\psi\Gamma_2\psi+\psi^*\Gamma_2^*\psi^*+\psi^*\Gamma_3\psi\,,
}
where we have assumed that $O_1$ and $O_2$ can be expressed as infinite series in powers of $\nabla^2$, so that spatial integrations by parts do not introduce minus signs. We have also defined
\eqa{
	\Gamma_1&\equiv O_1O_2^* ,\\
	\Gamma_2&\equiv \frac{1}{4}ma^3\left[O_2^2-O_1^2\mathcal{P}^2+\frac{3iH}{m}O_1O_2+\frac{i}{m}(O_1\dot{O_2}-O_2\dot{O_1})\right] ,\\
	\Gamma_3&\equiv -\frac{1}{2}ma^3\left[|O_2|^2+|O_1|^2\mathcal{P}^2+\frac{i}{m}(O_2\dot{O_1}^*-O_2^*\dot{O_1})\right]\,,
} 
and
\eq{
	\label{eq:P}
	\mathcal{P}\equiv \sqrt{1-\frac{\nabla^2}{m^2a^2}}\,.
}
From Eq.~\eqref{eq:lagnnlocal} we see that nonzero $\Gamma_2$ breaks $U(1)$ symmetry, which in turn causes nontrivial mode couplings, making the derivation of an EFT complicated. It would thus be plausible to eliminate $\Gamma_2$ by a suitable choice of $O_1$ and $O_2$. Let us define 
\eq{
	\mathcal{Q}\equiv O_1O_2^{-1}\,.
}
Then we can write
\eq{
	\label{Gamma2}
	\Gamma_2=\frac{1}{4}ma^3O_2^2\left[1-\mathcal{Q}^2\mathcal{P}^2+\frac{3iH}{m}\mathcal{Q}-\frac{i}{m}\dot{\mathcal{Q}}\right]\,.
}
Before considering an FLRW background, let us first consider the Minkowski case with $a (t) =1$ and $H (t) =0$. In this case we can find a solution for $\Gamma_2=0$ if we assume that $\mathcal{Q}$ is real (or pure imaginary), which results in two equations,
\eqa{
	1-\mathcal{Q}^2\mathcal{P}^2=0\,,\text{ and,}\quad 
	\dot{\mathcal{Q}}=0\,,	
}
which have a simultaneous simple solution, 
\eq{
	\mathcal{Q}=\mathcal{P}^{-1}\qquad\qquad\text{for Minkowski}\,.
}
Note that by setting $a (t) =1$, $\mathcal{P}$ becomes independent of time. Continuing with the Minkowski case, we can demand that $\psi$ be canonically normalized, which requires $\Gamma_1=1$, which in turn requires $O_1^*=O_2^{-1}$. Then we have
\eq{
	|O_1|=|O_2|^{-1}=\mathcal{P}^{-1/2}\,.
} 
Considering the nonrelativistic limit in Eq.~\eqref{eq:minlim}, a simple choice turns out to be 
\eq{
	O_1=O_2^*{}^{-1}=\mathcal{P}^{-1/2}e^{-imt}\, .
	\label{alphabetaMinkowski}
}
This reduces the general field redefinition to the nonlocal field redefinition introduced in Ref.~\cite{Namjoo:2017nia}. The Lagrangian then takes the form
\eq{
	\mathcal{L}=\frac{i}{2}\left(\dot{\psi}\psi^*-\dot{\psi}^*\psi\right)-m\psi^*(\mathcal{P}-1)\psi \quad\quad \text{for Minkowski}\,,
}
which has an explicit $U(1)$ symmetry. Note that the $U(1)$ symmetry will be broken in the presence of a self-interaction, indicating the violation of particle-number conservation. Note also that there is no oscillatory factor in this free theory, which is a major simplification, since mode coupling does not occur. This simplification, however, does not extend to the case of a self-interacting theory.

Returning to an FLRW background, we need to solve a differential equation for $\mathcal{Q}$ by requiring the terms in the brackets of Eq.~\eqref{Gamma2} to vanish, if we insist that $\Gamma_2=0$. However, we were unable to find a simple solution for that equation, and even requiring $\Gamma_2$ to be rather simple does not seem to simplify the derivation of an EFT. We therefore reverted to the local field redefinition, as introduced in Sec.~\ref{sec:redef} (see, however, Ref.~\cite{Friedrich:2019zic} for a generalization of the nonlocal operator to the case of curved geometry).

The nonlocal field redefinition of Eqs.~(\ref{phi_N}) and (\ref{alphabetaMinkowski}) in a Minkowski background enables us to derive the sound speed for density fluctuations, applicable for a wide range of momenta. To see this, first note that the equation of motion in this case is given by 
\ba 
\label{eq:sch_nonl}
i\dot \psi =m (\calP -1) \psi.
\ea 
In the low-momentum limit, we may expand the nonlocal operator $\calP$ to obtain the Schr\"odinger equation. However, Eq.~(\ref{eq:sch_nonl}) is exact (in Minkowski spacetime) and holds for arbitrary momentum. The Hamiltonian density in terms of the original field $\phi$ is given by
\ba 
\calH = \dfrac{1}{2} \dot \phi^2+ \dfrac{1}{2} (\nabla \phi)^2 +\dfrac{1}{2} m^2 \phi^2
\ea 
which, in terms of $\psi$, yields the Hamiltonian (representing the total energy of the system):
\ba
H =\int d^3x\left[ m \psi^*{\cal P} \psi  \right]=\int d^3x \left[m \vert{\cal P}^{1/2} \psi\vert^2  \right]\,,
\ea
where in the second equality we have performed an integration by parts. This implies that the energy density of the system can be written as $\rho = m \vert{\cal P}^{1/2} \psi\vert^2$. Inverting this relation suggests the relation $\psi = \calP^{-1/2}\left(\sqrt{\rho/m} \, e^{i\theta}\right)$, where $\theta$ is an arbitrary, but real, function of time and space. We assume that our Minkowski spacetime is filled with a homogeneous and isotropic condensate of particles, and then study small fluctuations around it (note also that we are ignoring the gravitational effects). Using the equation of motion for $\psi$, Eq.~\eqref{eq:sch_nonl}, at the background level we have 
\ba 
\dot{\bar \rho} =0, \qquad \dot{\bar \theta} =0.
\ea 
For small fluctuations, we find
\ba 
\dot \delta = 2m(\calP-1) \delta \theta \, , \qquad \dot{\delta \theta} = -\dfrac{m}{2} (\calP-1) \delta,
\ea 
where $\delta = \delta \rho/\bar \rho$. This yields
\ba 
\ddot \delta +m^2 (\calP-1)^2 \delta = 0.
\ea
Identifying this equation with the equation governing the propagation of a (massless) fluctuation with a nontrivial sound speed, namely 
\ba 
\ddot \delta -c_s^2 \nabla^2 \delta =0,
\ea 
and going to Fourier space, we conclude that the sound speed must take the form
\ba 
c_s^2 =\dfrac{m^2}{k^2} \left( \sqrt{1+\dfrac{k^2}{m^2}} -1 \right)^2 .
\label{eq:csfull}
\ea 
As emphasized above, we expect this relation to hold for arbitrary momentum (in Minkowski spacetime). In the small-momentum limit we obtain 
\ba  
c_s^2=\dfrac{k^2}{4m^2}-\dfrac{k^4}{8m^4}+\dots, \qquad {\text{for}} \quad k \ll m\, ,
\ea  
consistent with Eq.~\eqref{eq:eftcs}. Meanwhile, in the large-momentum limit, Eq.~(\ref{eq:csfull}) yields $c_s\to 1$, consistent with the speed of propagation of relativistic field fluctuations.

\section{Gauge transformation} \label{app:gauge}
In this appendix we investigate gauge transformations for fluctuations in our EFT. We mostly follow the conventions of Ref.~\cite{Weinberg:2008zzc}. The most general perturbed FLRW metric, considering only scalar perturbations, may be written
\eq{
\dd{s}^2=-(1+E)\dd{t}^2+2a(t)\p_iF\dd{t}\dd{x^i}+a(t)^2[(1+A)\delta_{ij}+\p_i\p_jB]\dd{x^i}\dd{x^j}\,.
}
For a general coordinate transformation of the form $x^\mu\to x^\mu+\delta x^\mu$, the metric fluctuations transform as
\eqa{
\Delta E=2\dot{\delta t}\, , \quad 
\Delta F=\frac{1}{a}\left( 2H\delta x-\dot{\delta x}-\delta t \right)\, , \quad 
\Delta A=2H\delta t\, , \quad 
\Delta B=-\frac{2}{a^2}\delta x\,,
\label{eq:gaugeEFAB}
} 
where we have defined $\delta x_0=\delta t$ and $\delta x_i=\p_i\delta x$. The scalar field $\phi$ also transforms as $\Delta \delta\phi=\dot{\bar{\phi}}\delta t$. However, we need a gauge transformation for $\dpsi$. From Eq.~\eqref{eq:psiphi} it is evident that $\delta \psi$ is \emph{not} a scalar under general coordinate transformations. We have
\eq{
\spl{
\Delta\dpsi=\psi'(x)-\psi(x)
=\psi'(x')-\psi(x)+\dot{\bpsi}\delta t\,,
}
}
where we have defined $x'{}^\mu=x^\mu+\delta x^\mu$.  As a precaution, note that in this appendix we use the prime to denote the variables in the new gauge, not to be confused with the rescaled time derivative in Sec.~\ref{sec:EFT}. In the new coordinates we can write
\eq{
\dot \phi'(x')=\pdv{\phi'(x')}{x'{}^0}=\pdv{x^\mu}{x'{}^0}\pdv{\phi(x)}{x^\mu}=\dot{\phi}(x)+\dot{\bar{\phi}}\dot{\delta t}\, .
}
As a result, from Eq.~\eqref{eq:psiphi} which defines $\psi$ we can show that
\eq{
\psi'(x')-\psi(x)=-im \, \bpsi \, \delta t+\frac{i}{\sqrt{2m}}e^{imt} \, \dot{\bar{\phi}} \, \dot{\delta t}\,.
}
Thus the gauge transformation for $\dpsi$ is
\eq{
\Delta\dpsi= \left( \dot{\bpsi}-im\bpsi \right) \delta t+\frac{1}{2}\left( \bpsi-e^{2imt}\bpsi^* \right)\dot{\delta t}\,.
\label{eq:Deltadpsigauge}
}
We can deduce the gauge transformation for each mode from these equations. To achieve this, we expand the coordinate transformation into different modes,
\eqa{
\delta t =\sum_\nu\delta t_\nu e^{imt}\, , \qquad 
\delta x =\sum_\nu\delta x_\nu e^{imt}\,.
}
Note that since, at this stage, $\delta t$ and $\delta x$ are arbitrary functions it is not necessarily the case that the slow modes (with $\nu = 0$) will dominate over the others. However, we will restrict ourselves to transformations for which such a hierarchy exists, as otherwise the gauge transformation may spoil the EFT construction for fluctuations. This is as a result of the fact that the field redefinition \eqref{eq:cantr1} breaks the general covariance and $\psi$ is not a scalar under general coordinate transformation. Substituting into the gauge transformation equations of Eq.~(\ref{eq:gaugeEFAB}), we find
\eqa{
&\Delta E_\nu=2\left( \dot{\delta t_\nu}+i\nu m\delta t_\nu \right)\,,\quad \Delta A_\nu=2H_\alpha \, \delta t_{\nu-\alpha}\,,\quad\Delta B_\nu=-2r_\alpha \,\delta x_{\nu-\alpha}\,,\\
&\Delta F_\nu=2q_\alpha H_\beta \, \delta x_{\nu-\alpha-\beta}-q_\alpha[\dot{\delta x}_{\nu-\alpha}+i(\nu-\alpha)m \, \delta x_{\nu-\alpha}]-q_\alpha\delta t_{\nu-\alpha}\,,
} 
where we have defined for simplicity $q \equiv 1/a(t)$. Similarly, from Eq.~(\ref{eq:Deltadpsigauge}), for $\dpsi$ we find
\eq{
\label{eq:dpsigauge}
\Delta\dpsi_\nu=\delta t_\alpha\bigg[\dot{\bpsi}_{\nu-\alpha}+im\left(\nu-\frac{\alpha}{2}-1\right)\bpsi_{\nu-\alpha}-im\frac{\alpha}{2}\bpsi^*_{\alpha+2-\nu}\bigg]+\frac{1}{2}\dot{\delta t}_\alpha \left( \bpsi_{\nu-\alpha}-\bpsi^*_{\alpha+2-\nu} \right)\,.
}
We can use these relations to find the equations governing the dynamics of fluctuations in a new gauge.

\subsection{Time-averaged comoving gauge}

In this paper, we have done our analyses mainly in the Newtonian gauge, defined by the line element in Eq.~\eqref{eq:ds2}. But the results of the previous section allow one to transform our EFT to other gauges. This procedure avoids tedious, from scratch, derivation of an EFT in a different gauge. As an example, one can consider the  \emph{time-averaged comoving} gauge, which is sometimes found convenient to work with. See, e.g., Ref.~\cite{Hwang:2009js,Poulin:2018dzj}, in which the derivation of the sound speed to leading order has been done in such a gauge. In our formalism, defining the time-averaged comoving gauge is fairly straightforward: We require to have $\delta u_{\text{eff}}=0$ in the new gauge.
Note that this is different from the standard comoving gauge in which $\du=-\delta\phi/\dot{\phi}=0$.  To find the relevant gauge transformation, $x'^\mu=x^\mu+\delta x^\mu$, we focus on the slow mode of the temporal part, i.e. we set $\delta t=\delta t_s$  and $\delta x=0$. By using the standard gauge transformation for the velocity potential $\Delta\delta u_{\text{eff}}=-\delta t$ and Eq.~\eqref{eq:dueff} for $\delta u_{\text{eff}}$ in the Newtonian gauge, we find that the appropriate transformation has to take the following form:
\eq{
\label{eq:dts}
\delta t_s=\frac{1}{2mi}\left(\frac{\dpsi_s}{\bar{\psi}_s}-\frac{\dpsi_s^*}{\bar{\psi}_s^*}\right)+\frac{3H_s}{4m^2}\left(\frac{\delta\psi_s}{\bar{\psi}_s}+\frac{\delta\psi_s^*}{\bar{\psi}_s^*}\right)+\frac{1}{2mi}\frac{\nabla^2}{4m^2a_s^2}\left(\frac{\delta\psi_s}{\bar{\psi}_s}-\frac{\delta\psi_s^*}{\bar{\psi}_s^*}\right)\,,
}
which leads to $\du'_{\text{eff}}=0$ in the new gauge. Further, by using the standard transformation of density and pressure fluctuations and their explicit expression in Newtonian gauge, Eqs.~\eqref{eq:drhoeff} and \eqref{eq:dpeff}, one can obtain their form in the new gauge and  show that $\dro'=\rho_{\text{eff}}\delta_{\text{eff}}$ and $\dP'_{\text{eff}}=c^2_{\text{eff}}\dro' _{\text{eff}}$ where $\delta_{\text{eff}}$ is gauge invariant and $c^2_{\text{eff}}$ is given by Eq.~\eqref{eq:eftcs}. 

As a consistency check,  we will try to confirm these results by applying the gauge transformation to the EFT in terms of $\dpsi_s$ and then obtain the equivalent fluid description, which we will see matches the above results. In the new gauge, as in Appendix~\ref{app:vis}, we write the metric as
\eq{
\label{eq:comov}
\dd{s}^2=-(1+2N)\dd{t}^2+2a(t)\p_i\sigma\dd{t}\dd{x^i}+a(t)^2(1+2\mathcal{R})\delta_{ij}\dd{x^i}\dd{x^j}\,,
}
where we can find the slow-mode parts of the new metric components in terms of  the old variables as
\eq{
\label{eq:newold1}
N_s=\Phi_s+\dot{\delta t}_s\,,\qquad\sigma_s=-\frac{\delta t_s}{a_s}\,,\qquad\mathcal{R}_s=-\Phi_s+H_s\delta t_s\,,
}
and for the field fluctuations we have
\eq{
\label{eq:newold2}
\dpsi'_s=\frac{\bar{\psi}_s}{2} \left[ \left( \frac{\delta\psi_s}{\bar{\psi}_s} + \frac{\delta\psi_s^*}{\bar{\psi}_s^*} \right) - \left( \frac{3iH_s}{m} + \frac{k^2}{2m^2a_s^2}\right) \frac{\dpsi_s^*}{\bar{\psi}_s^*} - \Phi_s\right]\,,
}
where we have used the gauge transformation in Eq.~\eqref{eq:dpsigauge} and the explicit form of $\delta t_s$ in Eq.~\eqref{eq:dts}. By inverting Eqs.~\eqref{eq:newold1} and \eqref{eq:newold2} we can write the old variables in terms of new ones as
\eqa{
&\dpsi_s=\dpsi'_s - ima_s \bar{\psi}_s \sigma_s + \frac{k^2}{4m^2a_s^2} \dpsi'_s - \frac{1}{2}\bar{\psi}_s \mathcal{R}_s-2a_sH_s \bar{\psi}_s\sigma_s\,,\\
&\Phi_s=-\mathcal{R}_s-a_sH_s\sigma_s\,.
}
Then, by using the effective equations \eqref{eq:eftsch}, \eqref{eq:eftphidot} and \eqref{eq:eftpoisson} we obtain the equations in the new gauge, which read
\eq{
\p_t(\dpsi'_s) - im \bar{\psi}_s \bigg( N_s+\mathcal{R}_s - \frac{5}{2}a_sH_s \sigma_s\bigg) + \frac{H_sk^2}{2m^2a_s^2}\dpsi'_s - H_s\bar{\psi}_s \bigg( \frac{3}{2}N_s+\frac{11}{4}\mathcal{R}_s + \frac{155}{16}a_sH_s\sigma_s\bigg)=0\,,
}
\eq{
\label{eq:poiscom}
\frac{\nabla^2\mathcal{R}_s}{a_s^2}+ \frac{H_s\nabla^2\sigma_s}{a_s^2} +\frac{1}{\Mpl^2}\left(m \bar{\psi}_s^*\dpsi'_s + \frac{3iH_s}{2} \bar{\psi}_s^* \dpsi'_s + \frac{\bar{\psi}_s^* k^2 \dpsi'_s}{4ma_s^2} + \frac{i|\bar{\psi}_s|^2 k^2 \sigma_s}{4a_s}\right)=0\,,
}
\eq{
\dot{\mathcal{R}}_s=H_sN_s\,,\qquad\dot{\sigma}_s+2H_s\sigma_s+\frac{1}{a_s}(N_s+\mathcal{R}_s)=0\,.
}
These are the effective equations in the new gauge.  Comparing Eq.~\eqref{eq:poiscom} with Eq.~\eqref{eq:comov2} for an imperfect fluid, we can identify the effective density fluctuation in this gauge as
\eq{
\dro' _{\text{eff}}=\rho_{\text{eff}}\left[2\frac{\dpsi'_s}{\bar{\psi}_s} + \frac{3iH_s \dpsi'_s}{m \bar{\psi}_s} + \frac{k^2 \dpsi'_s}{2m^2a_s^2 \bar{\psi}_s} + \frac{ik^2\sigma_s}{2ma_s}\right]\,.
}
As expected, this coincides with the definition $\dro'_{\text{eff}}=\rho_{\text{eff}}\delta_{\text{eff}}$ when we re-express the gauge invariant comoving overdensity in Eq.~\eqref{eq:deltaeffdef} in terms of variables in the new gauge. By further investigation and using the Euler equation in comoving gauge given in Eq.~\eqref{eq:comov3}, we can obtain the effective sound speed and coefficient of bulk viscosity which, not surprisingly, coincide with what we have already obtained in Newtonian gauge. This confirms the consistency of our results and, as a working example, shows the way that one can obtain the EFT and the fluid description in other gauges, making use of our results in the Newtonian gauge.

\bibliography{references} 
\bibliographystyle{JHEP}
\inputencoding{utf8}
\end{document}